\documentclass[11pt,a4paper]{article}
\usepackage{color}
\usepackage{amssymb}
\usepackage{amsmath}

\def\Im{\mathop{\rm Im}\nolimits}
\def\Re{\mathop{\rm Re}\nolimits}

\def\interior#1{\setbox1=\hbox{$#1$}\rlap{$#1$}\kern0.4\wd1\raise1.1\ht1%
\hbox{$\scriptstyle \circ$}}

\def\boxit#1#2{\setbox1=\hbox{\kern#1{#2}\kern#1}%
\dimen1=\ht1 \advance \dimen1 by #1 \dimen2=\dp1 \advance \dimen2 by #1
\setbox1=\hbox{\vrule height\dimen1 depth\dimen2\box1\vrule}%
\setbox1=\vbox{\hrule\box1\hrule}%
\advance \dimen1 by .4pt \ht1=\dimen1 \advance \dimen2 by .4pt \dp1=\dimen2
\box1\relax}
\def\endprf{\raise .5ex\hbox{\boxit{2pt}{\ }}}

\def\ifundefined#1{\expandafter\ifx\csname#1\endcsname\relax}

\def\beq{\begin{equation}}
\def\endq{\end{equation}}
\def\beqa{\begin{eqnarray}}
\def\endqa{\end{eqnarray}}

\let\UnmodifSec=\section
\renewcommand{\section}{\setcounter{equation}{0}\UnmodifSec}

\newtheorem{lemma}{Lemma}[section]

\def\Im{\mathop{\rm Im}\nolimits}
\def\Re{\mathop{\rm Re}\nolimits}

\def\interior#1{\setbox1=\hbox{$#1$}\rlap{$#1$}\kern0.4\wd1\raise1.1\ht1%
\hbox{$\scriptstyle \circ$}}

\def\boxit#1#2{\setbox1=\hbox{\kern#1{#2}\kern#1}%
\dimen1=\ht1 \advance \dimen1 by #1 \dimen2=\dp1 \advance \dimen2 by #1
\setbox1=\hbox{\vrule height\dimen1 depth\dimen2\box1\vrule}%
\setbox1=\vbox{\hrule\box1\hrule}%
\advance \dimen1 by .4pt \ht1=\dimen1 \advance \dimen2 by .4pt \dp1=\dimen2
\box1\relax}
\def\endprf{\raise .5ex\hbox{\boxit{2pt}{\ }}}

\def\ifundefined#1{\expandafter\ifx\csname#1\endcsname\relax}

\def\beq{\begin{equation}}
\def\endq{\end{equation}}
\def\beqa{\begin{eqnarray}}
\def\endqa{\end{eqnarray}}
\def\x{{u}}

\def\y{{v}}

\renewcommand{\j}{D}
\usepackage{graphicx}   

\def\a{m_1}
\def\b{m_2}
\def\c{m_3}
\def\aa{\zeta_1}
\def\bb{\zeta_2}

\def\cc{\zeta_3}
\def\intt{\int_0^\infty}
\allowdisplaybreaks[1]
\title{Banana integrals in configuration space}
\author{Sergio L. Cacciatori$^{(a,b)}$, Henri Epstein$^{(c)}$ and Ugo Moschella$^{(a,b,c)}$\\ \\ 
$^{(a)}$Disat, Universit\`a dell'Insubria, Como \\  $^{(b)}$INFN, Sezione di Milano, Italia\\ $^{(c)}$IHES, Bures-sur-Yvette, France\\}

\date{}
\begin{document}

\maketitle
\abstract{We reconsider the computation of banana integrals at different loops, by working in the configuration space, in any dimension.
We show how the 2-loop banana integral can be computed directly from the configuration space representation, without the need to resort to differential equations, 
and we include the analytic extension of the diagram in the space of complex masses. We also determine explicitly the $\varepsilon$ expansion of the two loop banana 
integrals, for $d=j-2\varepsilon$, $j=2,3,4$.}

\section{Introduction}
Ultraviolet divergences are an unavoidable crucial feature of Quantum Field theory (QFT). While infrared divergences, appearing in the presence of massless fields or in collinear beams of particles at high energies, can be cured by means of physical considerations \cite{Bloch:1937pw,Yennie,Weinberg:infrared}, ultraviolet divergences are more deeply related to the mathematical structure underlying the construction of the theory. 
They proliferate in perturbative formulations, requiring regularization at high momenta of the Feynman integrals and successive renormalization. When renormalization is controlled by a finite number of conditions, then fixing a finite number of external parameters,
possibly as functions of the energy scale, the theory is renormalizable. One of the most spectacular successes of QFT is the Standard Model of Particles, which, however, is not yet the final theory since it does not incorporate the gravitational field (in its full
quantum formulation) and is not free from problems (neutrino masses, quantization of the charges, $g-2$ for the muon, etc.). In the absence of a full nonperturbative formulation, the comparison of the very high-precision experiments performed nowadays requires to be able computing perturbative calculations at higher orders and expressing the results in the most possible compact and simple form.
Several efforts are done in this direction in recent years. A possibility is to reformulate the perturbative QFT in terms of positive Grassmannian geometry in a complexified momentum space. This strategy has led to the notion of Amplituhedron and its 
generalizations \cite{Arkani-Hamed:2013jha,Arkani-Hamed:2014dca,Arkani-Hamed:2020blm}, and has the advantage of potentially reducing the sum of the so-called kilo-Feynman to just the sum of few integrals. This line of research is developing rapidly and recently it has been shown that one of its realizations allows controlling ultraviolet and infrared divergences simultaneously \cite{Arkani-Hamed:2022cqe}. However, these methods do not apply in general yet but only work with particular theories. \\
Several other approaches are instead devoted to making the calculation of ``traditional'' Feynman integrals more efficient. One of these is the method of integration by parts, used to relate generic Feynman integrals of a given type to a small number of 
simpler integrals, called Master Integrals, explicitly known or easier to be computed analytically or numerically \cite{Chetyrkin:1981qh,Laporta:2003jz}. With the same method, one can compute differential equations for the Master Integrals, to be solved with
{specific}  boundary conditions. \\
{Another closely related method is} inspired by certain cohomological techniques originally developed in order {for a  deeper understanding} of hypergeometric integrals, see for example \cite{AomotoKita,MatsuGoto,MatsuTaka,Matsumoto}. The main idea is to interpret 
Feynman integrals as period integrals of some forms representing cohomological classes of a suitable twisted-cohomology. In this way, the set of Feynman integrals acquires a structure of linear space, endowed with a scalar product, given by the intersection 
product of the twisted cohomology \cite{MastroMiz,Mizera1,Mizera2,Weinzierl,Mastrolia}. Using this strategy is, therefore, easier to individuate a ``basis'' of master integrals, and then project any other integral in the same cohomology on the basis, or determine a 
Picard-Fuchs equation for the basis itself, by means of projections defined by the intersection product. Even this line of research is fast growing and has already {originated}  several developments and applications \cite{Frell1,Frell2,Frell3,Chestnov,MandalGasp,CaronPokr1,CaronPokr2,CheFreGaMaMa}. 

The same strategy applies not only to Feynman integrals but also to more general integrals involving special functions, typically appearing
in Quantum Mechanics or in Statistical Physics \cite{Cacciatori:2022mbi}. This suggests that the generic Feynman integrals can be tackled also in other representations rather than in the usual momentum space representation. 
An extended analysis including motivations for preferring the $x$-space to the momentum space can be found in \cite{Mishnyakov:2023wpd}.\footnote{We thank Alexei Morozov for putting this paper to our attention.} 
For example, this appears evident {when} looking for a relationship between Feynman integrals and the geometry of certain Calabi-Yau manifolds. In \cite{Bloch:2014qca}, the 3-bananas integral in two dimensions is written in the configuration space representation, therefore as an integral of the product of Macdonald functions, and specialized to the case of equal masses (normalized to 1) to find a differential Picard-Fuchs equation (w.r.t. $t=\sqrt K^2$, $K$ being the total momentum entering the banana diagram) whose solutions are used to compute the integral and then related to the motivic cohomology of a suitable $K3$ surface.
In \cite{Bonisch:2021yfw} this is generalized to the case of any $\ell$-banana integrals, still in two dimensions and equal masses, again starting from the representation in terms of Bessel functions. They are related to the motivic cohomology of specific Calabi-Yau manifolds. In \cite{Duhr:2022dxb}, the same strategy, once again in two spacetime dimensions, is extended to other classes of integrals. Banana integrals at any loop with all equal masses are studied in \cite{Pogel:2022vat}.\\
A review of the connection between vacuum banana integrals and integrals of products of Macdonald's functions can be found in \cite{Groote:2005ay}. In the context of conformal field theory, scalar 3-point functions are given by an integral of three Macdonald functions, determined in terms of Appell $F_4$ functions in \cite{Bzowski:2013sza}.  

In the present paper, we consider banana integrals up to two-loop order, with arbitrary masses, but with vanishing entering momentum. In \cite{Ford:1992pn} the two-loop banana integral for arbitrary masses and dimensions is explicitly solved by passing through the solutions of differential equations obtained from the momentum space representation of the integral. The solution is then used to compute the effective potential for the Standard Model of Particles up to two loops. It is clear that for more than two loops, the banana 
integrals are no more sufficient for computing the effective potential. However, our aim here is not to compute the effective potential at higher loops but rather to show the unexpected efficiency of working in the configuration space representation, in order to compute Feynman integrals or in finding differential equations they have to solve. The 0-momentum banana integrals thus allow us to compare our results with the ones in \cite{Ford:1992pn}.  
{It is worth mentioning that the x-space representation of Feynman integrals has already been used in literature for different aims than the present one. For example, in \cite{Broadhurst1,Broadhurst2,Zhou1,Zhou2,Broadhurst3,Zhou3} it played a crucial role in the determination of certain identities relating all equal masses $d=2$ sunrise integrals to modular integrals. In that case, {the restriction to equal masses} is crucial in order to factorize out the mass dependence and then map the problem to one of analytic number theory. However, we are {treating the general case where all the involved masses are not necessarily equal}, and in this case, it is not possible to obtain such a reduction. After putting $d=2$ and all masses equal to $1$ we do not get directly  {the same relations as in   \cite{Broadhurst1,Broadhurst2,Zhou1,Zhou2,Broadhurst3,Zhou3}}  but other {equivalent expressions. Our results might also be used to attempt a generalization of their formulas to different values of the dimension $d$. We leave this analysis for further research. On the other hand, it is interesting to notice that in that papers the differential equations are solved with methods avoiding the inversion of the Wronskian.\footnote{We acknowledge an anonymous referee for bringing this issue to our attention.} This also happens for all methods we used in the present paper.}}

In section \ref{flat}, we will warm up by computing the 1-loop banana integral
(the bubble integral), so reproducing the well-known standard result. 

In section \ref{3}, we show how the 2-loop banana integral can be computed directly from the configuration space representation, which is the integral of a product of Macdonald functions, without the need to resort to differential equations. 

In Section 4, we determine the analytic extension of the diagram in the space of complex masses.  

In Section 5, we provide very explicit formulas for the two-loop banana integral in dimension 2, 3 and 4.

In section \ref{5}, we recall the standard strategy of finding Picard-Fuchs equation for Feynman integrals, by reproducing the same equations used in \cite{Ford:1992pn} for the 2-loop banana integral. 

In Section \ref{6} we then show that the same equations are nothing but a manifestation of certain standard recursive relations among Macdonald functions, and the associated
Bessel-type second-order differential equation.

\section{The bubble and its momenta} \label{flat}
As a starter let us compute the bubble in $x$-space. In Euclidean Minkowski space the Schwinger function of a massive scalar field is proportional to a Macdonald function:
\begin{eqnarray}
 && G^d_m(x) =  \frac{1}{(2\pi)^{d }} \int \frac{e^{-ipx}}{{p^2+m^2}}  dp   = 
 \frac 1 {(2\pi)^{\frac d 2 }}   \left(\frac{r }{ m}\right)^{1-\frac{d}{2}}  K_{\frac{d}{2}-1}\left( m r \right), \ \ \ r=\sqrt{x^2},
\end{eqnarray}
where  $m$ is the mass of the field.  In $x$-space the bubble diagram  is represented by the following  integral:
\begin{align}
&
 \int G_{\a}(x)  G_{\b}(x)    dx   
= \frac{\left({ \a\b}\right)^{\frac{d}{2}-1}  }{2^{d-1}\pi ^{\frac d 2}\Gamma \left(\frac{d}{2}\right)}    
 \int_0^\infty     K_{\frac{d}{2}-1}\left( \a r \right)   K_{\frac{d}{2}-1}\left( \b r \right)  r dr      \label{aaa0}\\
  & = \frac{\Gamma \left(1-\frac{d}{2}\right)  \left({ \a\b}\right)^{\frac{d}{2}-1} }{2^{d}\pi ^{\frac d 2} }   
 \int_0^\infty   \left(  I_{\frac{d}{2}-1}\left( \a r \right)-  I_{1-\frac{d}{2}}\left( \a r \right)  \right)  K_{\frac{d}{2}-1}\left( \b r \right) r dr;    \label{aaa1}
\end{align}
in the last elementary but important step we used the identity
\begin{equation}K_\nu (z)=\frac{\Gamma (1-\nu ) \Gamma (\nu ) }{2}   (I_{-\nu}(z)-I_\nu(z)).
\label{bessel}
\end{equation}
The integral at the r.h.s. of (\ref{aaa0})   always converges at infinity.  On the other hand, since in an angle containing the positive real semiaxis in the complex $z$-plane 
\begin{equation} I_\nu(z) \sim \frac{e^z }{\sqrt{2\pi z}}, \ \ \ \ \ \ \  K_\nu(z) \sim {e^{-z} }{\sqrt{\frac{2\pi} z}},\end{equation}
the  integrals at the r.h.s. of Eq. (\ref{aaa1}) converge provided  $0< \a<\b$. 

By using the series representation 
\begin{align}
& I_{\nu }(z) =\sum _{n=0}^{\infty }{\frac {1}{n!\,\Gamma (n+\nu+1)}}   \left({\frac {z}{2}}\right)^{2n+\nu } \label{sser}
\end{align}
we may prove right away the well-known general formula 
\begin{align}
&\int_0^\infty   I_{\nu}\left( a r \right)  K_{\rho}\left( b r \right)  r  dr   =    \sum _{n=0}^{\infty } {\frac {(a/2)^{2n+\nu}}{n!\,\Gamma (n+\nu +1)}}  \int_0^\infty  r^{1+2n+\nu}     K_{\rho}\left( b r \right) dr \cr & 
=\frac{a^{\nu } \Gamma
   \left(\frac{\nu +\rho
   }{2}+1\right)
   \Gamma \left(\frac{\nu +\rho
   }{2}-1\right) \, _2F_1\left(\frac{\nu +\rho
   }{2}+1,\frac{\nu-\rho
   }{2}+1;\nu
   +1;\frac{a^2}{b^2}\right)}{ b^{\nu +2}\Gamma (\nu
   +1)}.
\end{align}
In the special case of interest to Quantum Field Theory $\rho =d/2-1$ and  $\nu =\pm (d/2-1) $;  the above formula immediately reduces to the textbook answer for the bubble:
\begin{eqnarray}
 \int G_{\a}(x)  G_{\b}(x)    dx = 
\frac { \Gamma \left(1-{\frac {d}{2}}\right)} {(4\pi)^{{\frac d 2}  }}  
 \frac{  m_2^{{d}-2}-m_1^{{d}-2} }{ m_1^2-m_2^2  } .
  \label{AAA0}
 \end{eqnarray}
 With the same simple steps, we may quickly find  the "moments" of the bubble as follows:
 \begin{align}
 & I_k(m_1,m_2,d)  = 
\frac{2\pi^{\frac d2}}{\Gamma\left(\frac d 2 \right)} \int_0^\infty r^k G_{\a}(r)  G_{\b}(r) \,  r^{d-1} dr  = 
\cr  
&=\frac{ \Gamma \left(\frac{k}{2}+1\right) \a^{d-k-4} \Gamma \left(\frac{4-d+k}{2}
   \right) \, _2F_1\left(\frac{k+2}{2},\frac{4-d+k}{2}
   ;\frac{4-d}{2};\frac{\b^2}{\a^2}\right)}{ 2^{d-k-
1}  \pi ^{\frac d2}\, (d-2) } +\cr &+ 
\frac{ \a^{-k-2} \b^{d-2} \Gamma \left(1-\frac{d}{2}\right) \Gamma
   \left(\frac{k}{2}+1\right) \Gamma \left(\frac{d+k}{2}\right) \,
   _2F_1\left(\frac{k+2}{2},\frac{d+k}{2};\frac{d}{2};\frac{\b^2}{\a^2}\right)}{ 2^{d-k} \pi ^{\frac{d}2}\Gamma
   \left(\frac{d}{2}\right)}.
  \label{momenta}
\end{align}
The unpleasant feature of the above formula is that the symmetry in the exchange of the masses $m_1$ and $m_2$ is not manifest. 

Always with the aim of explaining our methods in the simplest example, an explicitly symmetric formula is provided by the use of the Kallen-Lehmann representation (or linearization); we recall it for the reader's convenience:
\begin{equation}
G_{m_1}(x)\,G_{m_2}(x) = \int_{0}^\infty \,\rho(s, m_1, m_2)\, G_{\sqrt{s}}(x)\, ds \label{klformula}
\end{equation}
where
\begin{equation}
\rho(s, m_1, m_2)  
= { \left ( (s-(m_1+m_2)^2)(s-(m_1-m_2)^2) \right )^{d-3 \over 2}\over 2^{2d-3}\pi^{\frac{d-1} 2}
\Gamma \left ({d-1\over 2}\right )s^{d-2\over 2}}
\  \theta ((s-(m_1+m_2)^2).
\label{klweight}
\end{equation}
It follows that 
\begin{align}
 &I_k(m_1,m_2,d)  = 
\frac{2\pi^{\frac d2}}{\Gamma\left(\frac d 2 \right)} \int_0^\infty \int_0^\infty r^k \rho(s,\a,\b) \, G_{\sqrt{s}}(r) \,  r^{d-1} dr ds  
   \cr &=
  \frac{\Gamma \left(\frac{k}{2}+1\right) \Gamma \left(\frac{4-d+k}{2} \right)
    \, _2F_1\left(\frac{3-d}{2},\frac{4-d+k}{2} ;3-d;\frac{4 \a \b}{(\a+\b)^2}\right)}{\ 2^{d-k-1} \pi ^{\frac d 2}(d-2)\ (\a+\b)^{4-d+k}} +\cr 
    & +\frac{ \a^{d-2} \b^{d-2}  \Gamma \left(1-\frac{d}{2}\right) \Gamma
   \left(\frac{k}{2}+1\right) \Gamma \left(\frac{d+k}{2}\right)  \,
   _2F_1\left(\frac{d-1}{2},\frac{d+k}{2};d-1;\frac{4 \a \b}{(\a+\b)^2}\right)}{  2^{d-k} \pi ^{\frac d 2}\Gamma \left(\frac{d}{2}\right)\, (\a+\b)^{d+k}}. \label{momenta2}
\end{align}
Comparing Eqs. (\ref{momenta}) and (\ref{momenta2}) we deduce as a bonus the following remarkable identity:  for $a>b$
\begin{equation}
    \left(\frac{a} {a+b}\right)^{d+k} \, _2F_1\left(\frac{d-1}{2},\frac{d+k}{2};d-1;\frac{4 a b}{(a+b)^2}\right)= \, _2F_1\left(\frac{2+k}{2},\frac{d+k}{2};\frac{d}{2};\frac{b^2}{a^2}\right).
\end{equation}

\section{Two loops: the watermelon } \label{3}
In the previous simple example, we displayed the main ingredients of the calculation of a loop diagram in $x$-space:  the identity (\ref{bessel}), the series expansion (\ref{sser}), and the Kallen-Lehmann representation (\ref{klformula}).
We now exploit the same tools to compute the  harder two-loop watermelon:
\begin{align}
&I(m_1,m_2,m_3,d)= 
\int G_{\a}  G_{\b}   G_{\c}(x) dx = \frac{2\pi^{\frac d2}}{\Gamma\left(\frac d 2 \right)}\int G_{m_1}  G_{m_2}   G_{m_3}(r) r^{d-1} dr  \cr
&= \frac{2 \left({\a\b\c}\right)^{\frac{d}{2}-1} }{ 2^{\frac{3 d}{2}} \pi ^{d}\Gamma \left(\frac{d}{2}\right)}       
\intt  r^{2-\frac{d}{2}}  K_{\frac{d}{2}-1}\left( \a r \right)   K_{\frac{d}{2}-1}\left( \b r \right)    K_{\frac{d}{2}-1}\left( \c r \right) dr = \label{BB}
\\ &= 
\frac{\Gamma \left(1-\frac{d}{2}\right)^2 \Gamma
   \left(\frac{d}{2}\right)^2 }{4 }  
\sum_{\epsilon,\epsilon'=\pm} \epsilon\epsilon'R_{\epsilon\epsilon'}(\a,\b,\c\,d) \label{AAAbay}
\end{align}
where 
\begin{align} 
R_{\epsilon\epsilon'}(\a,\b,\c\,d)= \intt  r^{2-\frac{d}{2}}  I_{\epsilon (\frac{d}{2}-1)}\left(\a r \right)   I_{\epsilon'(\frac{d}{2}-1)}\left(\b r \right)    K_{\frac{d}{2}-1}\left( \c r \right) dr.   \label{AAAbay}
\end{align}
$I(m_1,m_2,m_3,d)$  actually depends on the squared masses. 

The integral at the r.h.s. of Eq. (\ref{BB})  always converges at infinity; it converges at $r=0$ in the strip
\begin{align}
 \Sigma= \{ d\in \mathbb C \, : \,  0< \Re d <3 \};
\end{align}
 it makes sense and defines a holomorphic function of the complex masses $m_1,m_2,m_3$, provided that $\Re m_j >0 $ for $j=1,2,3.$
The function $ I(m_1,m_2,m_3,d)$ at the l.h.s. coincides with the integral at the r.h.s. when the integral converges and is defined by analytic continuation otherwise. 

On the other hand,  the four integrals at the r.h.s. of (\ref{AAAbay}) converge at infinity only if $\Re \c>\Re \a+\Re \b$. Using Eq. (\ref{sser}) Bailey  \cite{bailey}  
proved in 1936 the following two elementary identifications:\footnote{We recall for the reader's convenience the definition of the Appell series of the first and of the fourth type 
\begin{align}
& F_{1}(a,b_{1},b_{2};c;x,y)=\sum _{m,n=0}^{\infty }{\frac {(a)_{m+n}(b_{1})_{m}(b_{2})_{n}}{(c)_{m+n}\,m!\,n!}}\,x^{m}y^{n}
=\frac {\Gamma(c)}{\Gamma(a)\Gamma(c-a)} \int_0^1 \frac {dt}t \frac {t^a (1-t)^{c-a-1}}{(1-xt)^{b_1} (1-yt)^{b_2}} 
\\
&F_{4}(a,b;c_{1},c_{2};x,y)=\sum _{m,n=0}^{\infty }{\frac {(a)_{m+n}(b)_{m+n}}{(c_{1})_{m}(c_{2})_{n}\,m!\,n!}}\,x^{m}y^{n}
\end{align}
}
\begin{align}
&\int_0^\infty  r^{\lambda-1}  I_{\mu}\left( a r \right)   I_{\nu}\left( b r \right)    K_{\rho}\left( c r \right) dr  = \cr = &
  \sum _{n=0}^{\infty }
\frac{2^{\lambda -2} a^{\mu } b^{\nu
   }
\Gamma\left(\frac{\lambda+\mu+\nu+\rho}2+n\right) \Gamma\left(\frac{\lambda+\mu+\nu-\rho}2+n\right) \,
  }{ c^{\lambda +\mu +\nu }\Gamma
   (\mu +1) \Gamma (n+1) \Gamma
   (n+\nu +1)} \times \cr
& \times\  {}_2F_1 \left(\frac{\lambda+\mu+\nu+\rho}2+n, \frac{\lambda+\mu+\nu-\rho}2+n; \mu+1; \frac {a^2}{c^2}\right) \frac {b^{2n}}{c^{2n}} \label{oop}= \\ 
=& 
 \frac{2^{\lambda-2} a^\mu b^\nu \Gamma\left(\frac{\lambda+\mu+\nu+\rho}2\right) \Gamma\left(\frac{\lambda+\mu+\nu-\rho}2\right)  F_4\left(\frac{\lambda+\mu+\nu+\rho}2, \frac{\lambda+\mu+\nu-\rho}2; \mu+1, \nu+1; \frac {a^2}{c^2}, \frac {b^2}{c^2}\right) }{  c^{\lambda+\mu+\nu}\     \Gamma(\mu+1)\Gamma(\nu+1) }  
\label{pp} 
\end{align}
valid for $\Re( \lambda +\mu+ \nu )> |\Re \rho | $ and $\Re (c\pm a\pm b) >0$.

\section{First derivation}
 When $m_3>m_1+m_2$, Eq. (\ref{pp}) allows the identification of  the watermelon with a sum of four Appell functions $F_4$:
 \begin{eqnarray}
 I(m_1,m_2,m_3,d)&= &
 (4 \pi) ^{1-d} m_3^{2 d-6} \frac{\Gamma (2-d)}{2 \sin
   \left(\frac{\pi  d}{2}\right) }
   F_4\left(3-d,2-\frac{d}{2},2-\frac{d}{2},2-\frac{d}{
   2},\frac{m_1^2}{m_3^2},\frac{m_2
   ^2}{m_3^2}\right) 
   \cr&+&{(4 \pi )^{-d}m_1^{d-2} m_2^{d-2}m_3^{-2}
   \Gamma \left(1-\frac{d}{2}\right)^2
   F_4\left(1,\frac{d}{2},\frac{d}{2},\frac{d}{2},\frac
   {m_1^2}{m_3^2},\frac{m_2^2}{m_3^2}\right)}
\label{AAA}\cr &-& (4 \pi )^{-d} m_1^{d-2}
  m_3^{d-4} \Gamma
   \left(1-\frac{d}{2}\right)^2
   F_4\left(1,2-\frac{d}{2},\frac{d}{2},2-\frac{d}{2},\
   \frac{m_1^2}{m_3^2},\frac{m_2^2}{
m_3^2}\right)  \cr &-& (4 \pi )^{-d} m_2^{d-2}
  m_3^{d-4} \Gamma
   \left(1-\frac{d}{2}\right)^2
   F_4\left(1,2-\frac{d}{2},2-\frac{d}{2},\frac{d}{2},\
   \frac{m_1^2}{m_3^2},\frac{m_2^2}{
m_3^2}\right). \cr \label{app3} &&\end{eqnarray} 
The above Appell functions may be reduced to the standard hypergeometric function by easy manipulations \footnote{Let us for instance  exhibit the few self-explanatory simple steps to compute the first term:
\begin{eqnarray}
&&  R_{++}(\a,\b,\c,d)= \intt  r^{2-\frac{d}{2}}  I_{\frac{d}{2}-1}\left( \a r \right)   I_{\frac{d}{2}-1}\left( \b r \right)    K_{\frac{d}{2}-1}\left( \c r \right) dr  \cr 
&&=  \sum _{n=0}^{\infty }\frac{2^{1-\frac{d}{2}}  \a^{\frac{d}{2}-1}  \b^{\frac{d}{2}-1} \,  _2F_1\left(n+1,\frac{d}{2}+n;\frac{d}{2};\frac{\a^2}{\c^2}\right)}{ c^{\frac{d}{2}+1}\Gamma \left(\frac{d}{2}\right)} \frac {\b^{2 n}}{ \c^{2 n}}
   \cr&&
\cr &&=  \sum _{n=0}^{\infty }
\frac{2^{1-\frac{d}{2}}
   \a^{\frac{d}{2}-1}
   \b^{\frac{d}{2}-1} \,
   _2F_1\left(n+1,-n;\frac{d}{2};\frac{{\a^2}}{{\a^2}-{\c^2}}\right)}{ \c^{\frac{d}{2}+1}\Gamma \left(\frac{d}{2}\right)} \frac {\b^{2 n}}{ \c^{2 n}}\left(1-\frac {\a^{2 }}{ \c^{2 }}\right)^{-n-1}
   \cr &&=  \frac{2^{1-\frac{d}{2}}
   \a^{\frac{d}{2}-1}
   \b^{\frac{d}{2}-1} \,
   }{ \c^{\frac{d}{2}+1}}  \sum _{n=0}^{\infty } \sum _{k=0}^{n } \frac{  \Gamma (k-n) \Gamma
   (k+n+1)   \left(\frac{\a^2}{\a^2-\c^2}\right)^k  \left( \frac {\b}{ \c}\right)^{2n}
 \left(\frac { \c^{2 } }{\c^2 -\a^{2 }}\right)^{n+1}}{\Gamma (k+1) \Gamma (-n)
   \Gamma (n+1) \Gamma
   \left(\frac{d}{2}+k\right)} 
    \cr && = \sum _{k=0}^{\infty } \sum _{n=k}^{\infty }\frac{2^{1-\frac{d}{2}}  \a^{\frac{d}{2}-1}  \b^{\frac{d}{2}-1} \,  }{ \c^{\frac{d}{2}+1}} \frac {\b^{2 n}}{ \c^{2 n}}\left(\frac{ \c^{2 }} {\c^2 - \a^{2 }}\right)^{n+1}\frac{   \left(\frac{\a^2}{\c^2-\a^2}\right)^   k }{\Gamma (k+1)  \Gamma  \left(\frac{d}{2}+k\right)}  \frac{ \Gamma (k+n+1)}{\Gamma   (-k+n+1)}
\cr&&
\cr&&
=\left( \frac{ \a
   \b }{ 2\c} \right)^{\frac{d}{2}-1}
     \frac{  \,
   _2F_1\left(\frac{1}{2},1;\frac{d}{2};\frac{4 \a^2
   \b^2}{\left(\c^2-\a^2-\b^2\right)^2}\right)}{ \Gamma
   \left(\frac{d}{2}\right) \left(\c^2-\a^2-\b^2\right)} . \label{steps}\end{eqnarray}
      The other terms are evaluated in a similar way.}
and there follows a simple symmetric formula for the watermelon:
\newpage \begin{eqnarray} I(m_1,m_2,m_3,d)  &=& \frac{
   \Gamma (2-d) \left(S(\a,\b,\c)\right)^{\frac{d}{2}-\frac{3}{2}}}{2^{2 d-1} \pi ^{d-1} \sin \left(\frac{\pi  d}{2}\right)}\cr 
 &+ & 
\frac{(\a\b)^{d-2} }{4^d \pi^{d-2}\sin ^2\left( 
   \frac{\pi d}{2}\right) \Gamma
   \left(\frac{d}{2}\right)^2}  
     \frac{  \,
   _2F_1\left(\frac{1}{2},1;\frac{d}{2};\frac{4 \a^2
   \b^2}{\left(\c^2-\a^2-\b^2\right)^2}\right)}{ \left(\c^2-\a^2-\b^2\right)} 
   \cr & & 
\cr & +& 
\frac{(\c\a)^{d-2} }{4^d \pi^{d-2}\sin ^2\left( 
   \frac{\pi d}{2}\right) \Gamma
   \left(\frac{d}{2}\right)^2}  
     \frac{  \,
   _2F_1\left(\frac{1}{2},1;\frac{d}{2};\frac{4 \c^2
   \a^2}{\left(\b^2-\c^2-\a^2\right)^2}\right)}{ \left(\b^2-\c^2-\a^2\right)} 
   \cr & & 
\cr &+ & 
\frac{(\b\c)^{d-2} }{4^d \pi^{d-2}\sin ^2\left( 
   \frac{\pi d}{2}\right)\Gamma
   \left(\frac{d}{2}\right)^2}  
     \frac{  \,
   _2F_1\left(\frac{1}{2},1;\frac{d}{2};\frac{4 \b^2
   \c^2}{\left(\a^2-\b^2-\c^2\right)^2}\right)}{ \left(\a^2-\b^2-\c^2\right)} 
    \label{bella formula}
\end{eqnarray}
where 
\begin{equation}
S(\a,\b,\c)=\a^4+\b^4+\c^4-2 \a^2 \b^2-2 \a^2 \c^2- 2 \b^2 \c^2 \label{311}
   \end{equation}
is the Symanzik polynomial.

The above formula is valid when one of the masses is bigger than the sum of the other two; this happens if and only if  the Symanzik polynomial is positive:
\begin{equation}
0<\frac{4 m_i^2m_j^2}{\left(m_k^2-m_i^2-m_j^2\right)^2}= 1-  \frac{S(m_1,m_2,m_3) }{\left(m_k^2-m_i^2-m_j^2\right)^2} <1, \ \ \ i\not= j   \not= k.
\end{equation}
The condition $S(m_1,m_2,m_3)>0$, in turn, implies that all the arguments of the hypergeometric functions on the r.h.s. of Eq. (\ref{bella formula}) are in the domain of convergence of the corresponding hypergeometric series and Eq. (\ref{bella formula}) can be taken at face value.

The Symanzik polynomial is positive in the particular case when one of the three masses is zero; in this case, the above formula simplifies to 
\begin{equation} I(\a,\b,0,d)  
 =
\frac{ 
   \Gamma (2-d)
   \left(\left(\a^2-\b^2\right)^2\right)^{\frac{d-3}{2}}}{2^{2d-1} \pi ^{d-1} \sin \left(\frac{\pi  d}{2}\right) }
   -
   \frac{ (\a \b)^{d-2}  \,
   _2F_1\left(\frac{1}{2},1;\frac{d}{2};\frac{4 \a^2
   \b^2}{\left(\a^2+\b^2\right)^2}\right)}{4^{d} \pi ^{d-2}\left(\a^2+\b^2\right)
   \sin
   ^2\left(\frac{\pi  d}{2}\right)\Gamma \left(\frac{d}{2}\right)^2}.
\end{equation}
Note also that a direct calculation would give an unsymmetrical result:
\begin{equation} 
I(\a,\b,0,d)  =
\frac{ 
   \Gamma (2-d)
   \left(\left(\a^2-\b^2\right)^2\right)^{\frac{d-3}{2}}}{2^{2d-1} \pi ^{d-1} \sin \left(\frac{\pi  d}{2}\right) }
   -
  \frac{    \Gamma
   \left(1-\frac{d}{2}\right) \,
   _2F_1\left(1,2-\frac{d}{2};\frac{d}{2};\frac{\b^2}{\a^2}\right)}{4^{ d}  \pi ^{d-1} \a^{2} (\a\b)^{2-d}\sin
   \left(\frac{\pi  d}{2}\right)\Gamma \left(\frac{d}{2}\right)}.
\end{equation}
 Comparing the above  equations we deduce the remarkable identity\footnote{This can be obtained from \cite{abramowitz}, by equating (15.3.16) to (15.3.17) and using
  \begin{align}
      a=1, \quad b=\frac d2-\frac 12, \quad z=\frac {4m_1m_2}{(m_1+m_2)^2}.
  \end{align}
  }
 \begin{eqnarray}
  {\left(1+\frac{\b^2}{\a^2}\right) \,
   _2F_1\left(1,2-\frac{d}{2};\frac{d}{2};\frac{\b^2}{\a^2}\right)}=
   { _2F_1\left(\frac{1}{2},1;\frac{d}{2};\frac{4
   \a^2 \b^2}{\left(\a^2+\b^2\right)^2}\right)} \label{49}
  \end{eqnarray}
valid at face value for $m_2<m_1$. 
This formula allows us to compare \eqref{bella formula} with the results in \cite{Bzowski:2015yxv}, see formulas (4.24) and (4.25). Starting from \eqref{BB}, after reversing the sign of all Bessel $K$ indices in the integral and multiplying by an appropriate power of the product of the masses, it reduces to (4.24) in \cite{Bzowski:2015yxv}. After using \eqref{49} we then get \eqref{bella formula}.\footnote{We thank Paul McFadden, Adam Bzowski, and Kostas Skenderis for pointing out these facts to us.}

When one of the masses is equal to the sum of the other two, then the Symanzik polynomial vanishes: all the arguments of the hypergeometric functions at the r.h.s. become equal to one while the argument of the last term vanishes.  We will compute the corresponding diagram below in Eq. (\ref{s=0}). 

When the  Symanzik polynomial is negative, or equivalently when each of the three masses is smaller than the sum of the other two (i.e.  when $m_1,m_2$, and $m_3$ are the sides of a triangle),  
none of the integrals at the r.h.s. of (\ref{AAAbay}) converge but a minor modification allows us to compute directly the diagram also in this circumstance. Suppose indeed that $\a<\b+\c$. Then
\begin{equation}
I(m_1,m_2,m_3,d) 
=   \frac{  \sum_{\epsilon=\pm} \epsilon\intt  r^{2-\frac{d}{2}}  I_{\epsilon (\frac{d}{2}-1)}\left(\a r \right)   K_{\frac{d}{2}-1}\left(\b r \right)    K_{\frac{d}{2}-1}\left( \c r \right) dr  }{ 2^{\frac{3 d}{2}-1}  \left({\a\b\c}\right)^{1-\frac{d}{2}}\pi ^{d-1} \sin\left(\frac{\pi  d}{2}\right)\Gamma \left(\frac{d}{2}\right)}  
 \label{AAAbaybis}
\end{equation} 
and now both the integrals at the r.h.s. of (\ref{AAAbaybis}) converge splendidly at infinity.

We now may insert in Eq. (\ref{AAAbaybis}) the series expansion (\ref{sser}), compute the integral using the formula for the moments (\ref{momenta}), and sum the resulting series following the same steps used to derive Eq. (\ref{steps}). This will produce the formula to be used when the Symanzik polynomial is negative. An alternative way makes use of the analyticity properties of the watermelon diagram in the three complex masses and is explained in the following Section \ref{4}.\footnote{As pointed out to us by P. McFadden, A. Bzowski, and K. Skenderis, the problem of analytically continuing in the masses also appears in the context of extracting flat-space scattering amplitudes from CFT correlators as discussed in \cite{Farrow:2018yni} and \cite{{Lipstein:2019mpu}}.  For this, one needs to analytically continue triple-K integrals to unphysical configurations where the sum of the momentum magnitudes vanishes.}


\subsection{Analytic continuation}
\label{4}
At first, we exploit the well-known  hypergeometric identity identity
\begin{eqnarray}    && _2F_1\left(\alpha,\beta ; \gamma;{z}\right)=  \frac{\Gamma (\gamma )\Gamma (\beta -\alpha  ) \, _2F_1\left(\alpha ,\alpha -\gamma +1;\alpha -\beta  +1;\frac{1}{z}\right)}{\Gamma (\beta ) \Gamma (\gamma  -\alpha )}  (-z)^{-\alpha } + \cr  && +\frac{\Gamma (\gamma ) \Gamma   (\alpha -\beta ) \, _2F_1\left(\beta ,\beta -\gamma  +1;-\alpha +\beta +1;\frac{1}{z}\right)}{\Gamma (\alpha ) \Gamma (\gamma -\beta )}  (-z)^{-\beta }, \ |\arg(-z)| <\pi \label{imp} \cr &&   \end{eqnarray}
to remodel our formula (\ref{bella formula}) in a way that may be used directly.  Let us, therefore, consider complex masses  $\aa,\bb,\cc$   such that none of the arguments of the hypergeometric functions at the r.h.s. of Eq. (\ref{bella formula}) is real. 
The identity  (\ref{imp})  has the virtue of disentangling the real and  imaginary parts of the various contributions in the limit when the arguments become real:  for instance, we have 
\begin{align} 
 R_{++}(\aa,\bb,\cc)   =&  \frac{ (\aa\bb)^{\frac{d}{2}-3}
   \left(-\aa^2-\bb^2+\cc^2\right) \,
   _2F_1\left(1,2-\frac{d}{2};\frac{3}{2};\frac{\left(-\aa^2-\bb^
   2+\cc^2\right)^2}{4 \aa^2 \bb^2}\right)}{ 2^\frac{d}{2} \cc^{\frac{d}{2}-1}\Gamma
   \left(\frac{d}{2}-1\right)}  +
   \cr + &\frac{ 2^{2-d} \sqrt{\pi }\ (2\aa\bb\cc)^{1-\frac{d}{2}}
 \left(-S(\aa,\bb,\cc) \right)^{\frac{d}{2}-\frac{3}{2}}}{\Gamma \left(\frac{d-1}{2}\right)  \left(-\aa^2-\bb^2+\cc^2\right)}  \sqrt{-{\left(-\aa^2-\bb^2+\cc^2\right)^2}},  \label{po1}
   \end{align}  
 and so on. Suppose then that $S(\a,\b,\c) <0$. There are three possibilities: 
\begin{enumerate}
\item The triangle is obtuse: the square of one of the masses is bigger than the sum of the squares of the other two, say $\c^2>\a^2  +\b^2$.

\item The triangle is acute-angled: no choice of the masses verifies the above inequality.

\item The triangle is right, say $\c^2=\a^2  +\b^2$.
\end{enumerate}

\vskip 10 pt 

\noindent $1. $ Suppose that $\c^2> \a^2  +\b^2$ and 
let $\cc= \c+i \epsilon$;  it is easily verified that 
\begin{eqnarray} 
&&\Im( -\left(\a^2-\b^2-\cc^2\right)^2)<0, \ \  \Im( -\left(-\a^2+\b^2-\cc^2\right)^2)<0, \cr  && \Im (-\left(-\a^2-\b^2+\cc^2\right)^2)<0, \ \  \Im S>0, 
\end{eqnarray}
and therefore 
\begin{eqnarray}
&& \Im (R_{++} -  R_{+-} - R_{-+}  + R_{--} )=  -i \frac{ 2^{2-d} \sqrt{\pi }\ (2\a\b\c)^{1-\frac{d}{2}}
 \left(-S(\a,\b,\c) \right)^{\frac{d}{2}-\frac{3}{2}}}{\Gamma \left(\frac{d-1}{2}\right)  }  \cr &&
   \cr && +2 i\frac{ 2^{2-d} \sqrt{\pi }\ (2\a\b\c)^{1-\frac{d}{2}}
 \left(-S(\a,\b,\c) \right)^{\frac{d-3}{2}}}{\Gamma \left(\frac{d-1}{2}\right)  } 
  \cr &&
 -i \frac{2^{2-d}  \sqrt{\pi }\ (2 \a \b \cc)^{\frac{1-d}{2}}    \left(-S(\a,\b,\c) \right)^{\frac{d-3}{2}}   \sin\left(\frac{d-3}{2}\pi\right) }{\cos\left(\frac{\pi  d}{2}\right)\Gamma
   \left(\frac{d-1}{2}\right)} 
= 0. \end{eqnarray}  

\noindent $2. $  Suppose that $m_i^2< m_j^2  +m_k^2$ for every $i\not = j \not = k$  and 
let $\aa= \a+i \epsilon$:  we have 
\begin{eqnarray} 
&&\Im( -\left(\aa^2-\b^2-\c^2\right)^2)>0, \ \  \Im( -\left(-\aa^2+\b^2-\c^2\right)^2)<0, \cr  && \Im (-\left(-\aa^2-\b^2+\c^2\right)^2)<0, \ \  \Im S<0. 
\end{eqnarray}
Again $ \Im (R_{++} -  R_{+-} - R_{-+}  + R_{--} )=  0$.
In both cases, the imaginary part of the sum of the various terms vanishes. This implies that it vanishes also in the limiting case  $\c^2= \a^2  +\b^2$.
Then, for $S(\a,\b,\c) <0$, the final result may be rewritten as follows:
\begin{align}
I(m_1,m_2,m_3,d)& =  -2^{1-2 d} \pi ^{1-d} \Gamma (2-d) (-S(\a,\b,\c))^{\frac{d-3}{2}}\cr 
&+ \frac{ (\a\b)^{d-4}
   \left(\a^2+\b^2-\c^2\right) \,
   _2F_1\left(1,2-\frac{d}{2};\frac{3}{2};M_{123}^2\right)}{4^{d} \pi ^{d-2}(\cos (\pi  d)-1)
   \Gamma \left(\frac{d}{2}-1\right) \Gamma
   \left(\frac{d}{2}\right)}
   \cr 
&+  \frac{ (\b\c)^{d-4}
   \left(-\a^2+\b^2+\c^2\right) \,
   _2F_1\left(1,2-\frac{d}{2};\frac{3}{2};M_{231}^2\right)}{4^{d} \pi ^{d-2}(\cos (\pi  d)-1)
   \Gamma \left(\frac{d}{2}-1\right) \Gamma
   \left(\frac{d}{2}\right)}  
   \cr 
&+  \frac{ (\a\c)^{d-4}
   \left(\a^2-\b^2+\c^2\right) \,
   _2F_1\left(1,2-\frac{d}{2};\frac{3}{2};{M_{312}^2}\right)}{4^{d} \pi ^{d-2}(\cos (\pi  d)-1)
   \Gamma \left(\frac{d}{2}-1\right) \Gamma
   \left(\frac{d}{2}\right)}.  \cr & \label{bf2}
\end{align}
where we defined 
\begin{equation} M_{ijk} = \left(\frac{m_i^2+m_j^2-m_k^2}{2 m_i m_j }\right)
\end{equation}
In the special important case where the three particles have the same mass the above formula reduces to
\begin{equation}
 I(m,m,m,d)=  \frac{  \frac 32 \ \Gamma
   \left(1-\frac{d}{2}\right)\Gamma \left(2-\frac{d}{2}\right)  \left(\,
   _2F_1\left(1,2-\frac{d}{2};\frac{3}{2};\frac{1}{4}\right)
   -\frac{\
   3^{\frac{d-5}{2}} \Gamma
   \left(\frac{3}{2}-\frac{d}{2}\right) \sqrt{\pi } }{ 2^{d-3}\Gamma \left(2-\frac{d}{2}\right)}\right)}{  (4 \pi) ^{d} }  \  m^{2 (d-3)} .  \end{equation}
By invoking the analyticity properties of the diagram in the complex masses
we also are able to evaluate the watermelon in the limiting case $S=0$, when one mass is equal to the sum of the other two:
\begin{align}
I(m_1,m_2,m_1+m_2,d)  = \frac{ \Gamma \left(1-\frac{d}{2}\right)
   \Gamma \left(2-\frac{d}{2}\right) \left( (\a+\b)^{d-3} \left(\b^3 \a^d+\a^3 \b^d\right)\right)-\a^d
   \b^d}{2^{2 d} \pi ^{d}(\a \b)^3
   (d-3) }
\label{s=0}.
\end{align}

\section{Evaluation near integer dimensions}
We will now consider the expansion in $\varepsilon$ near $\varepsilon\sim0$, for $$d=\j-2\varepsilon.$$ It is convenient to distinguish the dimensions according with their parity. 
\subsection{Odd dimensions} 
The odd-dimensional case is the easiest 
because
only the first line in formula \ref{bf2} has a (simple) pole at $\varepsilon=0$ due to the pole in the Gamma function
\begin{align}
    \Gamma(1-2k-2\varepsilon)=-\frac 1{(2k-1)!} \left( \frac 1{2\varepsilon} +\psi(2k) \right)+O(\varepsilon),
\end{align}
where 
$\psi$ is 
the logarithmic derivative of the gamma function.
Therefore
\begin{align}
    I(m_1,m_2,m_3,2k+1)=&\left( \frac 1{2\varepsilon}\psi(2k)-\frac 12 \log \frac {-S(m_1,m_2,m_3)}{16\pi^2}\right)\frac {(-S(m_1,m_2,m_3))^{k-1}}{2(4\pi)^{2k}(2k-1)!} \cr
    &-\frac {(m_1m_2)^{2k-3}(m_1^2+m_2^2-m_3^2) {}_2F_1(1,\frac 32-k;\frac 32;M_{123}^2)}{32 (4\pi)^{2k-1}\Gamma(k-1/2)\Gamma(k+1/2)} \cr
    &-\frac {(m_2m_3)^{2k-3}(m_2^2+m_3^2-m_1^2) {}_2F_1(1,\frac 32-k;\frac 32; M_{231}^2)}{32 (4\pi)^{2k-1}\Gamma(k-1/2)\Gamma(k+1/2)} \cr
    &-\frac {(m_3m_1)^{2k-3}(m_3^2+m_1^2-m_2^2) {}_2F_1(1,\frac 32-k;\frac 32;M_{312}^2)}{32 (4\pi)^{2k-1}\Gamma(k-1/2)\Gamma(k+1/2)} 
    \cr &+O(\varepsilon).
\end{align}
The hypergeometric functions appear in the form
\begin{align}
    {}_2F_1\left(1,\frac 32-k;\frac 32;z^2\right) =\frac 12 \int_0^1 \frac {dt}{\sqrt {1-t}} (1-z^2t)^{k-\frac 32},
\end{align}
and can be expressed in terms of elementary functions. For example for  $k=1$ (i.e $\j=3$) we have
\begin{align}
   & I(m_1,m_2,m_3,3-2\varepsilon)= \frac 1{32\pi^{2}} \frac 1{2\varepsilon} + \frac 1{32\pi^{2}} \left(  1 -\gamma-\frac 12 \log \frac {-S(m_1,m_2,m_3)}{16\pi^2}\right)
   \cr &-\frac{\tanh^{-1}\left(M_{123}\right)+\tanh^{-1}\left(M_{213}\right)+\tanh^{-1}\left(M_{312}\right)}{32 \pi ^2}+O(\varepsilon).
\end{align}

\subsection{Even dimensions. Case $\j=2$} 

\subsubsection{Triangular case}
We single out at first the case  $d=2$ which is non singular and start discussing the  triangular configuration where $S<0$. A Laurent series expansion of  Eq. (\ref{bf2}) shows that the coefficient of the possible diverging term vanishes thanks to the remarkable identity
\begin{equation}\arcsin \left(M_{123}\right)+\arcsin \left(M_{312}\right)+\arcsin \left(M_{231}\right)=\frac{\pi }{2};    \label{uiui}\end{equation}
Equation (\ref{uiui}) becomes obvious by using the identity
\begin{equation}
\arcsin M_{ijk} = \theta_{ijk} =  -i \log \left(\frac{1}{2m_im_j}\sqrt{-S}+i  M_{ijk}\right) \label{uiui0}
\end{equation}
A simple calculation shows that  
\begin{eqnarray}
&&  \left(\frac{1}{2m_1m_2}\sqrt{-S}+i  M_{123}\right) \left(\frac{1}{2m_2m_3}\sqrt{-S}+i  M_{231}\right)\left(\frac{1}{2m_3m_1}\sqrt{-S}+i  M_{312}\right)= \cr && =\exp({i\theta_{123}+ i\theta_{231}+i\theta_{312}})=\exp{\left(\frac{i\pi}{2}\right)}.\label{fase}
\end{eqnarray}
Therefore, the final result is completely fixed and there is no arbitrary regulatory mass floating around. 
The zero-th order term of the Laurent   series expansion includes a derivative of the hypergeometric function w.r.t the parameter $b$: 
\begin{align}
 &  I(m_1,m_2,m_3,2) = F(\a,\b,\c)+
 F(\c,\a,\b)+F(\b,\c,\a)
 \label{I2D} 
\end{align} 
where
\begin{align}
& F(\a,\b,\c) = \frac{M_{123}\, 
   {}_2\dot F_1\left(1,1,\frac{3}{2},M_{123}^2\right)}{16 \pi ^2 \a \b}
   +\frac{  \log \left(\frac{-S(\a,\b,\c)}{2 \, \a \b \c}\right)}{48 \pi  \sqrt{-S(\a,\b,\c)}} +
   \cr  &+\frac{\gamma  \left(\pi -6 \arcsin \left(M_{123}\right)\right)-6 \log \left(\frac{\a \b}{2
   \c}\right) \arcsin \left(M_{123}\right)}{48 \pi ^2 \sqrt{-S(\a,\b,\c)}}  
 \label{I2D2} 
\end{align} 
and we introduced the notation 
\begin{equation}
_2\dot{F}_1\left(a,x;c;z \right) = \frac{\partial }{\partial b }\left.  _2F_1\left(a,b;c;z\right)\right|_{b=x}=
-\int_0^1\frac{t^{a-1} \Gamma (c) (1-t)^{-a+c-1}  \log (1-t z)}{\Gamma (a) \Gamma
   (c-a)(1-t    z)^{x}}.
\end{equation}
A formula for
$
\dot{F}_1\left(1,1;\frac{3}{2};z\right)
$
in terms of the Euler-Spence dilogarithm function $\text{Li}_2$  may be obtained by  changing the integration variable  $t$ for $u = \sqrt {\frac {z(1-t)}{1-z}}$
and then by factorizing 
$(1+u^2)= (1+i u)(1-i u)$ whenever it appears. The final result is 
\begin{align}
    \dot{F}_1\left(1,1;\frac{3}{2};z\right)=& -\frac{\log \left(2 \sqrt{1-z}\right)}{\sqrt{(1-z) z}} \frac 1{2i}\left( \log \left(1+i \sqrt {\frac z{1-z}}\right)-
    \log \left(1-i \sqrt {\frac z{1-z}}\right)\right)\cr
    &+\frac 1{\sqrt{(1-z) z}} \frac 1{2i} \left( \text{Li}_2\left(\frac 12+\frac{i}{2} \sqrt{\frac{z}{1-z}}\right)-\text{Li}_2\left(\frac 12-\frac{i}{2} \sqrt{\frac{z}{1-z}}\right)\right) .
\end{align}
Note that in the second line  the imaginary part
of  Li$_2$ show up  only when $z$  real and such that  $0\leq z<1$. This happens exclusively in the triangular case. 

Inserting the above formula into Eq. (\ref{I2D})
we get 
\begin{align}
 &  I(m_1,m_2,m_3,2) = G(\a,\b,\c)+
 G(\c,\a,\b)+G(\b,\c,\a)
 \label{I2D3} 
\end{align} 
where
\begin{align}
& G(\a,\b,\c) = \frac{i \text{Li}_2\left(\frac{1}{2}-\frac{i
   \left(\a^2+\b^2-\c^2\right)}{2 \sqrt{-S}}\right)-i
   \text{Li}_2\left(\frac{1}{2}+ \frac{i \left(\a^2+\b^2-\c^2\right)}{2 \sqrt{-S}}\right)}{16 \pi ^2 \sqrt{-S}}
   \cr &-\frac{i \log (2 \c) \log \left(\frac{\sqrt{-S}+i
   \left(\a^2+\b^2-\c^2\right)}{2 \a \b}\right)}{8 \pi ^2 \sqrt{-S}} +\frac{\log \left(\frac{\sqrt{-S}}{2 \a \b \c}\right)}{48 \pi  \sqrt{-S}}.
\end{align}
Once more, only in the triangular case there appears the imaginary part
of the Euler-Spence dilogarithm Li$_2$; in this case the first line can be expressed in terms of the Bloch-Wigner\footnote{For general complex $z$ the definition of the Bloch-Wigner function $D(z)$ is 
\begin{align}
   D(z)= \Im {\rm Li}_2(z)+\arg(1-z)\log |z|,
\end{align}
$D(z)$  and has a number of interesting properties \cite{Zagier,Vanhove}.}  function $D$. 
By using Eq. (\ref{uiui0})
and by taking the cyclic sum we get
\begin{align}
    I(m_1,m_2,m_3,2) = \frac {1}{8\pi^2 \sqrt {-S}} D\left( \frac {\sqrt {-S} +i(m_1^2+m_2^2-m_3^2)}{2\sqrt{-S}}  \right)+cyc\{1,2,3\}.
\end{align}
It is interesting to compare our result to formula (3.10) of \cite{Vanhove}. 
Before doing this, it is worth noticing that such formula is correct as such only when the $\Delta$ appearing there is negative, namely, in the triangular case; when $z$ is real, one must replace the Bloch-Wigner function $D$ in formula (3.10) of \cite{Vanhove} with the second line of their formula (A.1). 

To make the comparison it is also necessary to remove a typo in formula (3.10) of \cite{Vanhove},
the correct coefficient being $4i\pi^2 \mu^2$. With this in mind, we get the interesting formula
\begin{align}
    2D\left( \frac {i\sqrt {-S} +m_1^2-m_2^2+m_3^2}{2m_1^2}  \right)=& D\left( \frac {\sqrt {-S} +i(m_1^2+m_2^2-m_3^2)}{2\sqrt{-S}}  \right)\cr +&D\left( \frac {\sqrt {-S} +i(m_1^2-m_2^2+m_3^2)}{2\sqrt{-S}}  \right)\cr +&D\left( \frac {\sqrt {-S} +i(-m_1^2+m_2^2+m_3^2)}{2\sqrt{-S}}  \right).
\end{align}
The r.h.s. is a manifestly symmetric function of the masses. The above relation  may be rewritten as a four term identity satisfied by the Bloch-Wigner function:
\begin{align}
    2D\left( \frac{2 x+i}{2 (x+y)}\right)=2D\left( \frac{2 y+i}{2 (x+y)}\right)= D\left( \frac 12 + i x \right)+D\left( \frac 12 + i y \right)+D\left(\frac{1}{2}+ \frac{ i(1-4 x y)}{4 (x+ y)}\right), \label{nostra}
\end{align}
valid for  $x$ and $y$ real. Indeed, this identity can be obtained from the Kummer formula (2) in \cite{Zagier}, by replacing 
\begin{align}
    z=\frac {1-z_2}{1-z_1-z_2},
\end{align}
where $z_1=\frac 12 +i x$, $z_2=\frac 12+iy$. Then, Kummer's formula gives
\begin{align}
    2D\left( \frac{2 y+i}{2 (x+y)}\right)=D\left( -\frac {1-z_1}{z_1}\right)+D\left( -\frac {1-z_2}{z_2}\right)+D\left( \frac {z_1 z_2}{z_1+z_2}\right).
\end{align}
By using the identity (\cite{Zagier}, (3))
\begin{align}
    D(w)=D\left( -\frac {1-w}{w}\right),
\end{align}
we get \eqref{nostra}.

\subsubsection{Non-triangular case}
Here $S>0$ and the starting point is Eq. (\ref{bella formula}); we suppose that $m_3>m_1+m_2$. There might exist first and second order poles in the result but, again, their coefficients vanish.
The zero-th order term of the Laurent   series  around $D=2$ now is given by 
\begin{align}
 &  I(m_1,m_2,m_3,2) = 
 \frac{\log (\a \b) \left(\log \left(\frac{\a \b}{16 \pi ^2}\right)+2 \gamma \right)}{8 \pi ^2
   \sqrt{S}}-
   \frac{\log (\a \c) \left(\log \left(\frac{\a \c}{16 \pi ^2}\right)+2 \gamma \right)}{8 \pi ^2
   \sqrt{S}} 
   \cr& 
  - \frac{\log ( \b\c) \left(\log \left(\frac{\b\c}{16 \pi ^2}\right)+2 \gamma \right)}{8 \pi ^2
   \sqrt{S}} +\frac{1}{48 \sqrt{S}}+\frac{\log (S) (\log (S)+4 \gamma -4 \log (4 \pi ))}{32 \pi ^2
   \sqrt{S}}+
   \cr &
   -\frac{{}_2F''_1\left(\frac{1}{2}
   ,1,1,\frac{4 \a^2
   \b^2}{\left(\a^2+\b^2-\c^2\right)^2}\right)+4 \left(\log
   \left(\frac{\a \b}{4 \pi }\right)+\gamma \right)
{{}_2F'_1\left(\frac{1}{2},1,1
   ,\frac{4 \a^2 \b^2}{\left(\a^2+\b^2-\c^2\right)^2}\right)}}{32
   \pi ^2 \left(\a^2+\b^2-\c^2\right)} 
   \cr &
   -\frac{{}_2F''_1\left(\frac{1}{2}
   ,1,1,\frac{4 \a^2
   \b^2}{\left(\a^2-\b^2+\c^2\right)^2}\right)+4 \left(\log
   \left(\frac{\a \c}{4 \pi }\right)+\gamma \right)
{{}_2F'_1\left(\frac{1}{2},1,1
   ,\frac{4 \a^2 \c^2}{\left(\a^2-\b^2-\c^2\right)^2}\right)}}{32
   \pi ^2 \left(\a^2-\b^2+\c^2\right)}
   \cr &
   -\frac{{}_2F''_1\left(\frac{1}{2}
   ,1,1,\frac{4 \b^2
   \c^2}{\left(-\a^2+\b^2+\c^2\right)^2}\right)+4 \left(\log
   \left(\frac{\a \b}{4 \pi }\right)+\gamma \right)
{{}_2F'_1\left(\frac{1}{2},1,1
   ,\frac{4 \a^2 \b^2}{\left(-\a^2+\b^2+\c^2\right)^2}\right)}}{32
   \pi ^2 \left(-\a^2+\b^2+\c^2\right)} \label{nontriangular}
   \end{align} 
%
where 
\begin{equation}
_2{F}'_1\left(a,b;x;z \right) = \frac{\partial }{\partial c }\left.  _2F_1\left(a,b;c;z\right)\right|_{c=x}, \ \ _2{F}''_1\left(a,b;x;z \right) = \frac{\partial^2 }{\partial c^2 }\left.  _2F_1\left(a,b;c;z\right)\right|_{c=x}
\end{equation}
For $(a,b,c)=(\frac 12, 1,1)$ they can be computed as follows.  First, we use the Kummer relation and assume\footnote{at the end, this condition can be relaxed by analytic continuation} $0\leq z<1/2$ to write
\begin{align}
    _2{F}_1\left(\frac 12,1;x;z \right)=&\frac 1{\sqrt{1-z}} _2{F}_1\left(\frac 12,x-1;x;-\frac z{1-z} \right)\cr
    =&\frac 1{\sqrt{1-z}} \frac {\Gamma(x)}{\sqrt \pi \Gamma\left(x-\frac 12\right)}\int_0^1 \frac {dt}{\sqrt t} \frac {(1-t)^{x-\frac 32}}{\left(1+\frac z{1-z}t \right)^{x-1}}. 
\end{align}
Taking the derivative w.r.t. $x$ in $x=1$ gives
\begin{align}
    _2{F}'_1\left(\frac 12,1;1;z \right) =-\frac 1{\sqrt{1 - z}} \frac 1\pi \int_0^1dt \frac 1{\sqrt {t(1-t)}} \log \left(1+\frac z{1-z}t \right).
\end{align}
After the first change of variable $t=s^2$ one can then rewrite the logarithm as a sum of two logarithms with argument linear in $s$. We are then left with the sum
of two integrals in which we take the changes $u=\sqrt{1-s^2}-is$ and $v=\sqrt{1-s^2}+is$. After simple manipulations, one gets that the sum of the integrals can be recast in an integral on the unit circle around the origin of the complex plane. Using the residue theorem gives
\begin{align}
    _2{F}'_1\left(\frac 12,1;1;z \right) =-\frac 2{\sqrt{1 - z}} \log \left( \frac 12 + \frac 12\frac 1{\sqrt{1 - z}}\right). \label{Fprimo}
\end{align}
Similar manipulation can be used for the second derivative, giving
\begin{align}
   _2{F}''_1\left(\frac 12,1;1;z \right) =& -\frac {\pi^2}3 \frac 1{\sqrt{1 - z}} + 
 \frac {8 \log2}{\sqrt{1 - z}} \log\left(\frac 2z (z - 1 + \sqrt{1 - z})\right) \cr &+ 
 \frac 2{\sqrt{1 - z}} \left({\rm Li}_2\left(  \frac 12 + \frac 12\frac 1{\sqrt{1 - z}}\right) - 
    {\rm Li}_2\left( \frac 12 - \frac 12\frac 1{\sqrt{1 - z}}\right)\right) \cr &+ 
 \frac 2{\sqrt{1 - z}} \log \left(\frac 12 + \frac 12\frac 1{\sqrt{1 - z}}\right) \log\left(-\frac {4 z}{1 - z}\right). \label{Fsecondo}
\end{align}
One sees that, for $0<z<1$ the argument of the logarithm in the second term at the RHS is negative and the argument of the first dilogarithm is greater than one. However, proceeding carefully, one sees that the two imaginary parts arising from those terms compensate each other and the formula stays real as the LHS is real. Actually the above equation can be rewritten\footnote{This can be made
 by using the identities
\begin{align}
    {\rm Li}_2(z)=&\frac {\pi^2}6 -{\rm Li}_2 (1-z)-\log z \log (1-z),\\
    \log \left(-\frac {4 z}{1 - z}\right)=&\log \left( \frac 12 - \frac 12\frac 1{\sqrt{1 - z}}\right)+\log \left( \frac 12 + \frac 12\frac 1{\sqrt{1 - z}}\right)
    +4 \log 2,\\
     \log\left(\frac 2z (z - 1 + \sqrt{1 - z})\right)=& -\log \left( \frac 12 + \frac 12\frac 1{\sqrt{1 - z}}\right).
\end{align}}
in a form which is free of the above shortcomings: 
\begin{align}
   _2{F}''_1\left(\frac 12,1;1;z \right) =& 
 \frac 2{\sqrt{1 - z}} \left(\log^2 \left(\frac 12 + \frac 12\frac 1{\sqrt{1 - z}}\right)-2{\rm Li}_2\left( \frac 12 - \frac 12\frac 1{\sqrt{1 - z}}\right) 
  \right). \label{FFsecondo}
\end{align}


By inserting Eqs. \eqref{Fprimo} and \eqref{FFsecondo} in \eqref{nontriangular} one gets an expression for the non-triangular case. The final result is not particularly illuminating and we do not write it down explicitly. 

However, there is an important point to be mentioned here. From \eqref{Fsecondo}, we
see that in the final formula the  Euler-Spence dilogarithms   
are evaluated at different 
real points. 
The Bloch-Wigner function, which is defined as the imaginary part of the dilogarithm plus a correction,
does not appear in the non-triangular case (and in general for complex masses).
 The expression (3.10) in \cite{Vanhove} thus contains the $D(z)$ function exclusively in the triangular case. We can further notice that the two cases (triangular and non) are related by analytic continuation in the complex plane, while
the Bloch-Wigner dilogarithm is not analytic in the complex plane, and, as such, not particularly relevant for quantum field theory.

\subsection{Even dimensions: $D=4$}
Here we discuss only the triangular  case $S<0$. For general even $D$, beyond the first-order pole from the first line, we have also in general second-order pole contribution 
from the remaining lines, since
$
    \cos (\pi  d)-1=-2\sin^2(\pi\varepsilon).
$   
We limit the discussion to $D=4$ which is obviously the physically relevant one but  has all the features of general case. 

A Laurent expansion gives
\begin{align}
 &  I(m_1,m_2,m_3,4+(d-4)) =  -\frac{\a^2+\b^2+\c^2}{128 \pi ^4 (d-4)^2} + \cr & +\frac{\left(3-2 \gamma +\log \left(16 \pi ^2\right)\right)
   \left(\a^2+\b^2+\c^2\right)}{256 \pi ^4 (d-4)}-\frac{\a^2 \log (\a)+\b^2
   \log (\b)+\c^2 \log (\c)}{64 \pi ^4 (d-4)} \cr & +\frac{\sqrt{-S} \left(\log \left(-\frac{S}{16
   \pi ^2}\right)+2 \gamma -3\right)}{512 \pi ^3}\cr & 
 -\left(\frac{\left(\a^2+\b^2-\c^2\right)
   \left({}_2\ddot F_1\left(1,0,\frac{3}{2},M_{123}^2\right)+\left(2-2 \log \left(\frac{\a^2 \b^2}{16 \pi ^2}\right)-4 \gamma \right)
   {}_2\dot F_1\left(1,0,\frac{3}{2},M_{123}^2\right)\right)}{1024 \pi ^4}+cyclic\right)
   \cr 
   & -\frac{1}{512 \pi ^4} \left((\a^2+\b^2+\c^2) \left(1+\frac {\pi^2}6\right) -2 \left( \a^2 L_{\a} + \b^2 L_{\b} +\c^2 L_{\c} \right)
   \right. \cr
   &+\a^2 L_{\a}^2 + \b^2 L_{\b}^2 +\c^2 L_{\c}^2+ (\a^2+\b^2-\c^2) L_{\a} L_{\b} \cr 
   &+ (\a^2-\b^2+\c^2) L_{\a} L_{\c} + (\a^2+\b^2-\c^2) L_{\b} L_{\c} \Big),
   \end{align} 
where 
\begin{align*}
    L_a:=\log \frac {a^2}{4\pi}+\gamma.
\end{align*}
Now, by using steps similar to those sketched in the two-dimensional case we get:
\begin{align}
 _2\dot F_1\left(1,0;\frac 3 2 ;z \right)  &=  2-2 \sqrt{\frac{1-z}{z}} \arctan
   \sqrt{\frac{z}{1-z}}
\\ 
    _2\ddot F_1\left(1,0;\frac 3 2 ;z \right) &= 4i\sqrt{\frac{1-z}{z}}\left({\rm Li}_2 \left( \frac {1+i\sqrt{\frac{1-z}{z}}}{1-i\sqrt{\frac{1-z}{z}}}\right)+\frac {\pi^2}{12}
    -\arctan^2 \sqrt{\frac{z}{1-z}}\right)
    \cr
     &+4\sqrt{\frac{1-z}{z}} \arctan \sqrt{\frac{z}{1-z}}(\log (4-4z)-2)+ 8.
\label{opop}\end{align}
In the triangular case  $z\in (0,1)$ and we can write
\begin{align*}
    \frac {1+i\sqrt{\frac{1-z}{z}}}{1-i\sqrt{\frac{1-z}{z}}}=e^{2i\psi}, \qquad \psi =\arctan \sqrt{\frac{1-z}{z}}=\frac \pi2-\arctan \sqrt{\frac z{1-z}}.
\end{align*}
The RHS of Eq. (\ref{opop}) is therefore real, as it can be checked by using the identity
\begin{align}
    {\rm Li}_2(e^{ix})=\frac {\pi^2}6-\frac \pi2 x+\frac 14 x^2+iCl_2(x),
\end{align}
valid for $x\in (-\pi,\pi)$, where $Cl_2$ is the Clausen function defined by
\begin{align}
   Cl_2(x)=\sum_{n=1}^\infty \frac {\sin (nx)}{n^2}=-\int_0^x \log \left| 2\sin \frac t2 \right|dt. 
\end{align}
More specifically, we can write
\begin{align}
 _2\dot F_1\left(1,0;\frac 3 2 ;\frac {(a^2+b^2-c^2)^2}{4a^2 b^2} \right)  &=  2+i \frac {\sqrt{-S}}{a^2+b^2-c^2} \log \frac {\sqrt{-S}+i(a^2+b^2-c^2)}{\sqrt{-S}-i(a^2+b^2-c^2)}
\\ 
    _2\ddot F_1\left(1,0;\frac 3 2 ;\frac {(a^2+b^2-c^2)^2}{4a^2 b^2} \right) &= 4i\frac {\sqrt{-S}}{a^2+b^2-c^2} \left({\rm Li}_2 
    \left( \frac {a^2+b^2-c^2+i\sqrt{-S}}{a^2+b^2-c^2-i\sqrt{-S}}\right) \right. \cr &
    \left. +\frac {\pi^2}{12}+\frac 14 \log^2 \frac {\sqrt{-S}+i(a^2+b^2-c^2)}{\sqrt{-S}-i(a^2+b^2-c^2)}\right)
    \cr
     -\frac {2i\sqrt{-S}}{a^2+b^2-c^2} \log& \frac {\sqrt{-S}+i(a^2+b^2-c^2)}{\sqrt{-S}-i(a^2+b^2-c^2)}\left(\log \frac {-S}{a^2b^2}-2\right)+ 8.
\label{opopop}
\end{align}
Replaced in $I(m_1,m_2,m_3,4+(d-4))$ it gives
\begin{align}
 &  I(m_1,m_2,m_3,4+(d-4)) =  -\frac{\a^2+\b^2+\c^2}{128 \pi ^4 (d-4)^2} + \cr 
 & +\frac{3 \left(\a^2+\b^2+\c^2\right)-2(\a^2 L_{\a}+\b^2 L_{\b}+\c^2 L_{\c})}{256 \pi ^4 (d-4)} \cr 
 & -\frac{1}{512 \pi ^4} \left((\a^2+\b^2+\c^2) \left(7+\frac {\pi^2}6\right) -6 \left( \a^2 L_{\a} + \b^2 L_{\b} +\c^2 L_{\c} \right)
   \right. \cr
 &+\a^2 L_{\a}^2 + \b^2 L_{\b}^2 +\c^2 L_{\c}^2+ (\a^2+\b^2-\c^2) L_{\a} L_{\b} \cr 
 &+ (\a^2-\b^2+\c^2) L_{\a} L_{\c} + (\a^2+\b^2-\c^2) L_{\b} L_{\c} \Big)\cr
 &-\frac {i\sqrt{-S}}{256 \pi^4}\left[ {\rm Li}_2 
    \left(\frac {\a^2+\b^2-\c^2+i\sqrt{-S}}{\a^2+\b^2-\c^2-i\sqrt{-S}}\right) \right.\cr
  &  \left.+\frac {\pi^2}{12}+\frac 14 \log^2 \frac {\sqrt{-S}+i(\a^2+\b^2-\c^2)}{\sqrt{-S}-i(\a^2+\b^2-\c^2)} +cyc\{1,2,3\}\right]. 
\end{align} 
By using the above relation with the Clausen function, with $\psi\equiv\frac \pi2-\theta_{ijk}$, after introducing the Lobachevsky function
\begin{align}
    L(\theta)=-\int_0^{\theta} dx\ \log |\cos x|,
\end{align}
so that
\begin{align}
    L(\theta_{ijk})=\theta_{ijk} \log 2-\frac 12 Cl_2(\pi-2\theta_{ijk}),
\end{align}
and finally using \eqref{fase}, we get
\begin{align}
   &i  \left[ {\rm Li}_2 
    \left(\frac {\a^2+\b^2-\c^2+i\sqrt{-S}}{\a^2+\b^2-\c^2-i\sqrt{-S}}\right) 
   +\frac {\pi^2}{12}+\frac 14 \log^2 \frac {\sqrt{-S}+i(\a^2+\b^2-\c^2)}{\sqrt{-S}-i(\a^2+\b^2-\c^2)}\right.\cr 
   &+cyc\{1,2,3\}\Big]=-\left(Cl_2(\pi-2\theta_{123})+Cl_2(\pi-2\theta_{231})+Cl_2(\pi-2\theta_{312})  \right)\cr
   &=\left(L(\theta_{123})+L(\theta_{231})+L(\theta_{312})-\frac \pi2 \log 2\right).
\end{align}
Therefore,
\begin{align}
 &  I(m_1,m_2,m_3,4+(d-4)) =  -\frac{\a^2+\b^2+\c^2}{128 \pi ^4 (d-4)^2} + \cr 
 & +\frac{3 \left(\a^2+\b^2+\c^2\right)-2(\a^2 L_{\a}+\b^2 L_{\b}+\c^2 L_{\c})}{256 \pi ^4 (d-4)} \cr 
 & -\frac{1}{512 \pi ^4} \left((\a^2+\b^2+\c^2) \left(7+\frac {\pi^2}6\right) -6 \left( \a^2 L_{\a} + \b^2 L_{\b} +\c^2 L_{\c} \right)
   \right. \cr
 &+\a^2 L_{\a}^2 + \b^2 L_{\b}^2 +\c^2 L_{\c}^2+ (\a^2+\b^2-\c^2) L_{\a} L_{\b} \cr 
 &+ (\a^2-\b^2+\c^2) L_{\a} L_{\c} + (\a^2+\b^2-\c^2) L_{\b} L_{\c} \Big)\cr
 &+\frac {\sqrt{-S}}{128 \pi^4}\left(L(\theta_{123})+L(\theta_{231})+L(\theta_{312})-\frac \pi2 \log 2\right). 
\end{align} 
This coincides with formula (4.20) in \cite{Ford:1992pn}.


\subsection{The 1-loop sunset}
A little adaption of the above results provides a formula for the sunset at 1-loop:
this is nothing but the Fourier transform  of the product of two Schwinger functions
\begin{align}
& Sun(k,m_2,m_3,d)= 
 \int e^{ikx} G_{\b}(x)  G_{\c}(x)    dx=
 \cr &= 
  \frac{1}{(2\pi)^{\frac d2}}  \left(\frac{k}{\b\c}\right)^{1-\frac{d}{2}} \int  r^{2-\frac{d}{2}} J_{\frac{d}{2}-1}\left( k r \right) K_{\frac{d}{2}-1}\left( \b r \right)K_{\frac{d}{2}-1}\left( \c r \right) dr .
 \cr
\end{align}
By using Eqs. (\ref{steps})  we immediately get
\begin{align}
 Sun =  \frac{ \c^{d-2} \,
   _2F_1\left(\frac{1}{2},1;\frac{d}{2};-\frac{4 \c^2
   k^2}{\left(-\b^2+\c^2-k^2\right)^2}\right)}{2^{d} \pi ^{\frac{d}{2}-1}\sin
   \left(\frac{\pi  d}{2}\right) \Gamma
   \left(\frac{d}{2}\right) \left(\b^2-\c^2+k^2\right)}-\frac{ \b^{d-2}  \,
   _2F_1\left(\frac{1}{2},1;\frac{d}{2};-\frac{4 \b^2
   k^2}{\left(\b^2-\c^2-k^2\right)^2}\right)}{2^{d} \pi ^{\frac{d}{2}-1} \sin
   \left(\frac{\pi  d}{2}\right)\Gamma
   \left(\frac{d}{2}\right) \left(\b^2-\c^2-k^2\right)}
\end{align}
valid for  values of $k$ small enough. In the limit where the two masses are equal the above formula reduces to 
\begin{align}
& Sun(k,m,m,d)= 
\frac{ 2\,  m^{d-2} \Gamma
   \left(1-\frac{d}{2}\right) \,
   _2F_1\left(\frac{1}{2},1;\frac{d}{2};-\frac{4
  m^2}{k^2}\right)}{ 2^{d} \pi ^{\frac d 2}k^2}.
   \end{align}
Similarly, by using Eqs. (\ref{po1}) and similar, we get 
\begin{align}
 Sun = &  -\frac{ (d-2)\Gamma
   \left(1-\frac{d}{2}\right) \left(-\b^2+\c^2+k^2\right) \,
   _2F_1\left(1,2-\frac{d}{2};\frac{3}{2};-\frac{\left(\b^2-\c
   ^2-k^2\right)^2}{4 \b^2 k^2}\right)}{ 2^{d+2}  \pi ^{\frac d 2} \b^{4-d}   k^2}  \cr
   & -\frac{(d-2) \Gamma
   \left(1-\frac{d}{2}\right) \left(\b^2-\c^2+k^2\right) \,
   _2F_1\left(1,2-\frac{d}{2};\frac{3}{2};-\frac{\left(\b^2-\c
   ^2+k^2\right)^2}{4 \c^2 k^2}\right)}{ 2^{d+2} \pi ^{\frac d 2} \c^{4-d} k^2}
\end{align}
valid for  values of $k$ large enough.
In the limit where the two masses are equal the above formula reduces to 
\begin{align}
& Sun(k,m,m,d)= 
\frac{ \left(4
   m^2+k^2\right)^2 \,
   _2F_1\left(1,2-\frac{d}{2};-\frac{1}{2};-\frac{k^2}{4
   m^2}\right)+4  (d-6) m^2 k^2 -16 m^4}{ 2^{d} \pi ^{\frac{d}{2}-1} (d-5) (d-3)  \sin
   \left(\frac{\pi  d}{2}\right)     \Gamma \left(\frac{d}{2}-1\right)m^{4-d} k^4
}.   \end{align}
\subsection{Three Loops: an example}
Here the general task would be to compute

\begin{align}
I(m_1,m_2,m_3,m_4,d)= & 
\int G_{\a}(x)  G_{\b}(x)   G_{\c}(x) G_{m_4} (x)dx \cr 
=& \frac { \omega_{d} }{(2\pi)^{ {2d}  }}   \left(\frac{1 }{\a\b\c m_4}\right)^{1-\frac{d}{2}}    K(m_1,m_2,m_3,m_4d)  
\end{align}
where
\begin{equation}
 K(m_1,m_2,m_3,m_4,d)= 
\intt  r^{3-d}  K_{\frac{d}{2}-1}\left( \a r \right)   K_{\frac{d}{2}-1}\left( \b r \right)    K_{\frac{d}{2}-1}\left( \c r \right) K_{\frac{d}{2}-1}\left( m_4 r \right)  dr  \label{BBBB}
\end{equation}
The same methods applied before allow to derive
formulae in  special cases. For instance 
\begin{eqnarray}
&&  K(m,m,M,M,d)= 
\intt  r^{3-d}  K_{\frac{d}{2}-1}\left(m r \right)   K_{\frac{d}{2}-1}\left( m r \right)    K_{\frac{d}{2}-1}\left( M r \right) K_{\frac{d}{2}-1}\left( M r \right)  dr  =\cr&&
-\frac{\pi ^{3/2} 2^{d-6} m^{2 d-6} M^{2-d} \Gamma
   \left(4-\frac{3 d}{2}\right) \Gamma (3-d) \Gamma \left(\frac{d}{2}-1\right) \,
   _2F_1\left(4-\frac{3 d}{2},\frac{3-d}{2};\frac{7}{2}-d;\frac{M^2}{m^2}\right)}{\sin \left(\frac{\pi  d}{2}\right) \Gamma
   \left(\frac{7}{2}-d\right)}  \cr&&
   -\frac{\pi ^2 2^{-d} m^{d-4}  \Gamma
   \left(1-\frac{d}{2}\right) \,
   _3F_2\left(\frac{1}{2},1,3-d;\frac{5}{2}-\frac{d}{2},\frac{d}{2};\frac{M^2}{m^2}\right
   )}{(d-3) \sin ^2\left(\frac{\pi  d}{2}\right)\Gamma \left(\frac{d}{2}-1\right)} \cr&&
   +\frac{\pi  2^{-d-1} M^{d-2}  \Gamma
   \left(1-\frac{d}{2}\right) \Gamma \left(2-\frac{d}{2}\right) \,
   _3F_2\left(1,2-\frac{d}{2},\frac{d}{2}-\frac{1}{2};\frac{3}{2},d-1;\frac{M^2}{m^2}\right)}{m^2 \sin \left(\frac{\pi  d}{2}\right)}
   \end{eqnarray}
 The most interesting case $M=m$ follows.
\section{PDE's for banana integrals: a summary} \label{5}
Here we reconsider the method of PDEs applied to banana integrals in $x$-space.

The usual way to tackle the calculation of Feynman's diagrams is to start from their momentum space representations. For the watermelon, this is
\begin{align}
F (\x,\y,z,d)  =&\frac{1}{(2\pi)^{3d }} \int \frac{e^{-ikx}}{{k^2+\x}}   \frac{e^{-iqx}}{{q^2+\y}} \frac{e^{-ipx}}{{p^2+z}}  dkdqdpdx  \label{b2}\\
=& \frac{1}{(2\pi)^{2d }} \int \frac {dq dk }{(k^2+\x)(q^2+\y)((q+k)^2+z)}  =   I(\sqrt{\x} ,\sqrt{\y},\sqrt{z},d).
\end{align} 
The trick to deduce a partial differential equation (PDE) for $F (\x,\y,z,d)$ makes use of Stokes' theorem as, for instance, in the following example:
\begin{align}
0=&   \frac{1}{(2\pi)^{2d }}  \int dq dk\, \frac{\partial} {\partial k^\mu} \, \frac {k^\mu}{(k^2+\x)(q^2+\y)((q+k)^2+z)}  \cr 
=&  F  (d-3) -  2\x  \frac{\partial F }{\partial \x}-   (\x -\y+z) \frac{\partial F }{\partial z} - J(\x,\y,z),  \label{a1}
\end{align} 
where
\begin{equation}
J(\x,\y,z)= -
\frac{\Gamma
   \left(1-\frac{d}{2}\right) \Gamma
   \left(2-\frac{d}{2}\right)  \left( \x^{\frac d 2-1}-
   \y^{\frac d 2-1}\right) z^{\frac{d}{2}-2}}{ (4 \pi )^{d} }
\end{equation}
satisfies the identity
\begin{equation}
J(\x,\y,z) z + J(z, \x,\y) \y +J(\y,z,\x) \x=0.
\end{equation}
Interchanging $\x$ and $\y$  in  Eq. (\ref{a1}) we get a second independent equation:
 \begin{eqnarray}
 F  (d-3) -  2\y  \frac{\partial F }{\partial \y}-   (\y -\x+z) \frac{\partial F }{\partial z} - J(\x,\y,z)=0. \label{aa4}
\end{eqnarray} 
By summing and subtracting  Eqs. (\ref{a1})  and (\ref{aa4})  they are replaced by\footnote{Eqs (\ref{aa3}) and (\ref{aaa4}) follow directly from the vanishing of the  integrals
\begin{equation}
 \int dq dk\,\left( \frac{\partial}{\partial k^\mu }\, \frac {k^\mu}{(k^2+\x)(q^2+\y)((q+k)^2+z)}\pm \frac{ \partial}{\partial q^\mu }\, \frac {q^\mu}{(k^2+\x)(q^2+\y)((q+k)^2+z)} \right) = 0
\end{equation} }
 \begin{eqnarray}
&&  \x  \frac{\partial F }{\partial \x}  -  \y  \frac{\partial F }{\partial \y}  +(\x-\y)  \frac{\partial F }{\partial z}+   J(\x,\y,z)=0 ,\label{aa3} \\
&& (d-3) F-  \x \frac{\partial F }{\partial \x} -  \y \frac{\partial F }{\partial \y} -  z \frac{\partial F }{\partial z} =0. \label{aaa4}
\end{eqnarray} 
A third independent  equation may be obtained by interchanging the roles of $\y$ and $z$ in Eq. (\ref{aa3})
 \begin{eqnarray} \x  \frac{\partial F }{\partial \x}  -  z  \frac{\partial F }{\partial z}  +(\x-z)  \frac{\partial F }{\partial \y}+   J(\x,z,\y) =0.\label{a5} 
\end{eqnarray} 
The remaining equation obtained by interchanging $\x$ and $z$  is not independent of the other two; 
however, the sum of the three equations obtained in this way  
coincide with the  symmetric equation solved in \cite{Ford:1992pn} to derive a formula for  the watermelon:
 \begin{eqnarray}
 (\x-\y)  \frac{\partial F }{\partial z}  +(\y-z)  \frac{\partial F }{\partial \x}+ (z-\x)  \frac{\partial F }{\partial \y}+  J(\x,\y,z)+ J(z,\x,\y)+ J(\y,z,\x)= 0. \label{eng}  \end{eqnarray} 
Because of their independence and their linearity, Eqs. (\ref{aa3}), (\ref{aaa4}) and (\ref{a5})  may be used to disentangle the partial derivatives of $F$:
 \begin{eqnarray}
\frac{\partial F }{\partial \x}= \frac{(d-3)(\x-\y-z) F(\x,\y,z) +2 J(\x,\y,z) z+ J(\y,z,\x) (\x-\y+z)}{\x^2+\y^2+z^2-2 \x
   \y-2 \x z-2 \y z} ; \label{pd}
\end{eqnarray} 
the other derivatives ${\partial F }/{\partial \y}$ and ${\partial F }/{\partial z}$  are obtained by cyclic permutations of the variables $\x,\y$ and $z$. 

For example, the derivative of the watermelon w.r.t. say $m_1^2$ takes the following form:
  \begin{align}
 \frac{\partial I }{\partial \a^2}=&  I(\a,\b,\c) \frac{\partial \log S^\frac{d-3}2 }{\partial \a^2}  +\frac{ \Gamma \left(1-\frac{d}{2}\right)
   \Gamma \left(2-\frac{d}{2}\right)}{ (4 \pi )^{d} }    \times \cr \times&{ \frac{ 2 \a^4 \b^d\c^d- \a^d \b^2\c^d
   \left(\a^2+\b^2-\c^2\right)- \a^d \b^d \c^2
   \left(\a^2-\b^2+\c^2\right)}{\a^4 \b^2 \c^2 \ S(\a,\b,\c)} }. \label{popo} 
   \end{align} 
Another noticeable  symmetric equation where the Symanzik polynomial explicitly appears:
\begin{align}
\frac{\partial F }{\partial \x}  +   \frac{\partial F }{\partial \y}  +\frac{\partial F }{\partial z}    =&- \frac{(d-3)(\x+\y+z) F(\x,\y,z) }{\x^2+\y^2+z^2-2 \x
   \y-2 \x z-2 \y z} + \cr   +&\frac{J(\x,\y,z) (\y-\x)+ J(\y,z,\x) (z-\y)+J(z,\x,\y) (\x-z)}{\x^2+\y^2+z^2-2 \x   \y-2 \x z-2 \y z} .
\end{align} 
The conclusion is that the set of equations \eqref{aa4}, \eqref{aa3}, and \eqref{a5} fully characterizes the banana integral we are analyzing; eq.  \eqref{aa4} is just reflecting the homogeneity of the integral.
This is due to the fact that the set of equations obtained by differentiating the integral w.r.t. the masses is a $D$-module of dimension 3. 

The $D$-module for the two-loop banana integral (or the sunrise) is well-known as well as the equations discussed 
in the present section. They are equivalent to the equations obtained in \cite{Remiddi} or the ones of \cite{Ford:1992pn}. { Then,} one usually looks for a Gr\"obner basis of the module; { here we do not, since we are investigating a way to determine an optimal expression for the explicit solution of the integral rather than a way for constructing a basis
for such equations. In passing we remark that the equations exhibited in \cite{Ford:1992pn} look to us more advantageous than the ones in \cite{Remiddi}} since they lead directly to an expression of 
the two-loop banana integral { manifestly} symmetric in the masses;  this simplifies the task of the analytic continuation to the physical range of interest. While the analysis in \cite{Remiddi} is complete from the point of view of characterizing and determining
a set of Master Integrals, { it } would make the derivations of the results of \cite{Ford:1992pn} much harder.

\section{PDE's for loop diagrams: a fresh look in position space} \label{6}
In this section we propose a way to { derive} convenient equations { suitable for providing} explicit solutions of the banana integrals with arbitrary masses and dimensions. We do this directly in $x$-space. 
The main idea is that in order to look for more symmetric representations of the integral one should look for symmetric expressions of the nonhomogeneous terms. 
Inspired by \cite{Ford:1992pn}, we look for a way
to reproduce their equations for the two-loop banana integral, in a manner that can be extended to higher loop cases.\\

Let us focus again on Eq. (\ref{a1}).
 A useful modification is to apply Stokes' trick to the r.h.s. of Eq. (\ref{b2}) as follows:
\begin{eqnarray}
\frac{1}{(2\pi)^{3d }} \int \frac{\partial}{\partial k^\mu}   \frac{ k^\mu  e^{-ikx}}{{k^2+\x}}   \frac{e^{-iqx}}{{q^2+\y}} \frac{e^{-ipx}}{{p^2+z}}   dkdqdpdx  =0.
\end{eqnarray}
At this point we may perform first  the integration over the $k$ variable  (and leave the integration over $x$ at the last step): formally we get
\begin{eqnarray}
 \frac{1}{(2\pi)^{d }} \int \frac{\partial}{\partial k^\mu}  \frac{ k^\mu  e^{-ikx}}{{k^2+m^2}}  dk
=(d -2)G^d_{m} (x) - 2m^2 \partial_{m^2} G^d_{m} (x) -2\pi r^2 G^{d+2}_{m} (x) = 0.
\end{eqnarray}
In terms of MacDonald functions the above identity is indeed well-known (Eq. (\ref{nuK})) and it amounts to  
\begin{eqnarray}
m^{\frac d 2} r^{2-\frac{d}{2}} K_{-\frac{d}{2}}(m r) -  (d -2)m^{\frac{d}{2}-1} r^{1-\frac{d}{2}} K_{1-\frac{d}{2}}(mr) - m^{\frac d 2} r^{2-\frac{d}{2}}
   K_{2-\frac{d}{2}}(m r)=0. \label{rec}
 \end{eqnarray}
The conclusion is summarized in the following 

\begin{lemma} The partial differential equation (\ref{a1}) is  equivalent to the recurrence relation (\ref{rec}) among Macdonald functions.
\end{lemma}

The point that we want to make now is that indeed all the PDEs described in Sect. \ref{5} arise from   the modified Bessel equation and  the known recursion relations for the Macdonald functions.\footnote{We list them here for reference \cite{bateman2}:
\begin{eqnarray}
&& z^2   \partial^2_z K_\nu(z)+z \partial_z K_\nu (z)-\left( {z^2} + {\nu^2}\right) K_\nu(z)=0, \label{eqK}\\
&&   2 \partial_z K_\nu (z)+K_{\nu-1}(z)+K_{\nu+1}(z) =0 , \label{derK}\\
&& K_{\nu-1}(z)-K_{\nu+1}(z)+ 2  {\nu}{z^{-1}} K_\nu (z) = 0,\label{nuK} \\
&&\partial_r (r^{\nu } K_{\nu }(m r))+m r^{\nu } K_{\nu -1}(m r)=0. \label{recc}
\end{eqnarray}}

Before proceeding it is worthwhile to stress that our method might work also in curved spacetimes where a global linear momentum space is not available; furthermore, it may also be used to obtain rapidly new equations also in flat space as we will do at the end of this chapter.

Let us start by exhibiting a few basic formulae. 
\begin{eqnarray}
\frac{\partial  G^d_m(r)}{\partial m^2} =  -
 \frac 1 {2(2\pi)^{\frac d 2 }}   \left(\frac{r }{ m}\right)^{\frac{4-d}{2}}  K_{\frac{d-4}{2}}\left( m r \right) = - \frac 1 {4\pi}    G^{d-2}_m(r). \label{demg}
 \end{eqnarray}
where we used both  \eqref{derK} and \eqref{nuK}. Similarly
\begin{eqnarray}
\frac{\partial  G^d_m(r)}{\partial r} = - \frac 1 {(2\pi)^{\frac d 2 }} m^{\frac d 2} r^{1-\frac{d}{2}}
   K_{\frac{d}{2}}(m r)   = -2\pi r \, G^{d+2}(r). \label{derg}
\end{eqnarray}
Together they give 
\begin{align}
 \partial_r\partial_{m^2}G_m^d(r)=& \frac r{2} G_m^{d}(r). \label{dermg}
\end{align}
Furthermore
\begin{eqnarray}
\frac{\partial^2  G^d_m(r)}{\partial r^2} 
=m^2 G_m^d(r) +2\pi(d-1) G_m^{d+2}(r). \label{der2g}
\end{eqnarray}
Finally, it is useful to rewrite also the recurrence (\ref{recc}) in terms of the Schwinger functions: 
\begin{eqnarray}
 (d-2) G^d_m(r) - 2\pi r^2 G_m^{d+2}(r) +\frac{ m^2}{2\pi} G_m^{d-2}(r) =0 .\label{reccg}
\end{eqnarray}

Now let us proceed with the derivation of two other PDEs by working only in $x$-space.
Using  Eqs. (\ref{demg}), \eqref{derg} and (\ref{reccg}) we get (the argument $r$ in $G$ is omitted):
\begin{eqnarray}
&& m_1^2 \frac{\partial I }{\partial m_1^2} = -\frac {  m_1^2 \omega_d} {4\pi}\int_0^\infty r^{d-1} G_{m_1}^{d-2}  G_{m_2}^{d}  G_{m_3}^{d} dr \cr &&
 = \frac {\omega_d}{2} \left(  d-2\right) \int_0^\infty r^{d-1} G_{m_1}^{d}  G_{m_2}^{d} G_{m_3}^{d} dr -  {\pi \omega_d}\int_0^\infty r^{d+1}   G_{m_1}^{d+2} G_{m_2}^{d}  G_{m_3}^{d} dr \cr 
&& =  \left( \frac d2-1\right) I +\frac {\omega_d}2 \int_0^\infty r^d  (\partial_r G_{m_1}^{d}) G_{m_2}^{d} G_{m_3}^{d} dr .\label{questa}
\end{eqnarray}
Symmetrization in the masses gives 
\begin{align}
 \left( m_1^2 \partial_{m_1^2}+m_2^2 \partial_{m_2^2}+m_3^2 \partial_{m_3^2}\right)I=3 \left( \frac d2-1\right) I +\frac {\omega_d}2  \int_0^\infty r^d  \partial_r \left( G_{m_1}^{d} G_{m_2}^{d}  G_{m_3}^{d} \right) dr.
\end{align}
When  $0<Re(d)<3$ the  boundary term obtained by partial integration vanishes and we recover Eq. (\ref{aaa4}): 
\begin{align}
 \left( m_1^2 \partial_{m_1^2}+m_2^2 \partial_{m_2^2}+m_3^2 \partial_{m_3^2}\right)I=(d-3) I.  \label{pop}
\end{align}
Finally,  the analyticity properties of the function $I(\a,\b,\c,d)$ guarantee that Eq. (\ref{pop}) holds without restriction on the dimension $d$. 

In the following example, the role of boundary terms at $r=0$ may be better appreciated.
By interchanging the role of $\a$ and $\b$ in Eq. \eqref{questa} we get
\begin{eqnarray}
&&  \left( m_1^2 \partial_{m_1^2}-m_2^2 \partial_{m_2^2} \right)I= \frac {\omega_d}2 \int_0^\infty r^d G_{m_3}^{d} \left(G_{m_2}^{d} \partial_r G_{m_1}^{d}  -G_{m_1}^{d}  \partial_rG_{m_2}^{d}   \right) dr\cr 
& =& \omega_d \partial_{m_3^2} \int_0^\infty r^{d-1}  \partial_rG_{m_3}^{d} \left(G_{m_2}^{d}  \partial_r G_{m_1}^{d}  -G_{m_1}^{d}  \partial_r G_{m_2}^{d}   \right) dr 
\cr &=&b.t. - \omega_d \partial_{m_3^2} \int_0^\infty (d-1)r^{d-2} G_{m_3}^{d} \left(G_{m_2}^{d} \partial_r G_{m_1}^{d}  -G_{m_1}^{d}  \partial_r G_{m_2}^{d}   \right) dr \cr
 & &-\omega_d \partial_{m_3^2} \int_0^\infty r^{d-1} G_{m_3}^{d} \left(G_{m_2}^{d}  \partial_r^2 G_{m_1}^{d}  -G_{m_1}^{d}  \partial_r^2 G_{m_2}^{d}   \right) dr. 
\end{eqnarray}
A comment is in order concerning the second step, where we used Eq. (\ref{dermg}): in that equation, the derivative w.r.t. $m^2$ cancels a term that close to $r= 0$ behaves differently than at the l.h.s.; when the derivative is taken outside the integral the convergence of the latter gets worst and it only works for  $0<Re(d)<2$. In the third step, we integrated by parts 
denoting by $b.t.$  the boundary terms. By inserting  Eq. \eqref{derg} in the first line and Eq. \eqref{der2g} in the second we get the following equation: 
\begin{align}
  \left( m_1^2 \partial_{m_1^2}-m_2^2 \partial_{m_2^2} \right)I=& b.t. - (m_1^2-m_2^2)  \partial_{m_3^2} I.
\end{align}
There remains the evaluation of the boundary terms. For for $0<Re(d)<2$ the leading terms of the Schwinger function at $r\sim 0$ are
\begin{eqnarray}
G_{m}^{d}(r)\simeq \frac {\Gamma
   \left(1-\frac{d}{2}\right)}{ (4 \pi )^{\frac d 2}} m^{d-2}  +\frac{ \Gamma \left(\frac{d}{2}-1\right)}{ 4\pi^{\frac d 2}} r^{2-d},
 \ \ \   \partial_r G_{m}^{d}(r)\simeq 
   -\frac{ \Gamma \left(\frac{d}{2}\right)}{2 \pi^{\frac d 2}} r^{1-d}
\end{eqnarray}
so that 
\begin{align}
 b.t. =& \omega_d \partial_{m_3^2} \int_0^\infty  \frac d{dr} \left[r^{d-1} G_{m_3}^{d} \left( G_{m_2}^{d} \frac {d}{dr}G_{m_1}^{d}  -G_{m_1}^{d} \frac {d}{dr}G_{m_2}^{d}   \right) \right]dr\cr
=& \omega_d \lim_{r\to0}  \left[r^{d-1} \partial_{m_3^2}G_{m_3}^{d} \left( G_{m_1}^{d} \partial_r G_{m_2}^{d}  -G_{m_2}^{d}\partial_r G_{m_1}^{d}   \right)\right]\cr
=&- \frac {\partial G_{m_3}^{d}}{\partial {m_3^2}} (0) \left( G_{m_1}^{d}(0) -G_{m_2}^{d}(0)   \right) =-J(m_1^2,m_2^2,m_3^2).
\end{align}
All in all, we recover Eq. (\ref{aa3}):
\begin{align}
  \left( m_1^2 \partial_{m_1^2}-m_2^2 \partial_{m_2^2} \right)I+ (m_1^2-m_2^2)  \partial_{m_3^2} I +J(m_1^2,m_2^2,m_3^2)=0.
\end{align}

\section{Conclusions and perspectives}
By considering the explicit example of zero-momentum banana integrals with arbitrary masses and in any dimensions, we have investigated the potentiality of using the configuration space representation to compute the Feynman integrals.
In particular, we studied very explicitly the case of two loops. After expressing the banana integral as an integral of the product of three Macdonald functions, we have used two strategies in order to compute them. On one hand, by means of certain Bailey's 
formulas known in the mathematical literature, we have expressed the banana integral as a combination of $F_4$ Apple's functions. On the other hand, we have shown that quite simple manipulations of the series expansion of the modified Bessel functions
it is possible to rewrite the banana integral as a combination of (much simpler) ${}_2F_1$ hypergeometric functions, manifestly symmetric in the masses, a result directly comparable to the one in \cite{Ford:1992pn} but obtained in an elementary way, without recurring to the solution of differential equations. Moreover, we studied the analytic extension of such solutions thus providing the necessary formula for all possible physical cases. Interestingly, by comparing the two different expressions, we get an interesting relation between certain combinations of $F_4$ Appel's functions and corresponding combinations of Gauss' hypergeometric functions.\\
We have then investigated the Picard-Fuchs equations associated with the banana integrals by showing that they can be obtained in a quite elementary way from the configuration space representation: they are simply a direct consequence of the standard recursive relations satisfied by the modified Bessel functions and the modified Bessel equation. 
\\
There are several possible perspectives we want to consider for future work. First, it could be interesting to generalize our construction to the case of non-zero external momentum, for applications to scattering theory. Another possibility is to consider more general zero momentum loop integrals as, e.g., the ones necessary to compute the 3-loops effective potential for the standard model. Further, by combining with the methods in \cite{Bonisch:2021yfw,Duhr:2022dxb}, it may be that one can identify a more general relation with cohomological structures and with the intersection theory methods.\\
Finally, and perhaps more interesting for justifying the configuration space representation, is to try applying the same philosophy to the case of quantum field theory on a curved background, where the momentum representation is not available. Some of these topics are under consideration for further work. 
\newpage

\appendix

\section{Another formula for the watermelon and a corollary}
Another interesting formula for the two-loop watermelon  which involves only one Appell function $F_1$, may be obtained by using the Kallen-Lehmann representation: 
\begin{align}
 &I(m_1,m_2,m_3,d)= \int dx\int_0^\infty \rho(s,m_1,m_2) G_{m_3}(x)G_{\sqrt{s}}(x) ds     \cr  &=
\int_{(\b+\c)^2}^\infty \frac{   \left(\frac {\sqrt s^{2-d}}{\a^{2-d}}-1\right)
   \left(s-(\b-\c)^2\right)^{\frac{d-3}{2}} \left(s-(\b+\c)^2\right)^{\frac{d-3}{2}}}{  2^{2 d-1} \pi ^{d-1}\sin \left(\frac{\pi  d}{2}\right) \left(s-\c^2\right)
   \Gamma (d-1)}  ds=  \\
    &=-\frac{  
   \left((\a-\b)^2-\c^2\right)^{\frac{d-3}{2}} \left((\a+\b)^2-\c^2\right)^{\frac{d-3}{2}}}{2^{2 d-1} \pi ^{d-2}\sin \left(\frac{\pi  d} {2}\right) \sin (\pi  d)\Gamma (d-1)} +\cr & 
-\frac{ \a^{d-2} \b^{d-2} \Gamma
   \left(1-\frac{d}{2}\right) \, _2F_1\left(1,\frac{d-1}{2};d-1;\frac{4 \a \b}{(\a+\b)^2-\c^2}\right)}{4^{d} \pi ^{d-1}  \sin \left(\frac{\pi  d}{2}\right) \Gamma
   \left(\frac{d}{2}\right)\left((\a+\b)^2-\c^2\right)} +\cr 
  &+ \frac{ \c^{d-2}  \Gamma \left(2-\frac{d}{2}\right)
   (\a+\b)^{d-4}
   F_1\left(2-\frac{d}{2};\frac{3-d}{2},1;\frac{3}{2};\frac{(\a-\b)^2}{(\a+\b)^2},\frac{\c^2}{(\a+\b)^2}\right)}{2^{3 d-4} \pi ^{d-1}\sin \left(\frac{\pi  d}{2}\right)\Gamma \left(\frac{d}{2}\right)}. \label{bellafiglia}  
\end{align}
The above formula is valid when $\c^2 < \a^2 + \b^2$.

By comparing Eqs.   (\ref{bella formula}) and (\ref{bellafiglia}) we obtain a far from obvious summation formula for  the Appell series $F_1$ appearing in Eq.  (\ref{bellafiglia}):  
\begin{lemma}
\begin{align}
   & F_1\left(\frac{4-d}{2};\frac{3-d}{2},1;\frac{3}{2};\frac{(a-b)^2}{(a+b)^2},\frac{c^2}{(a+b)^2}\right)= \cr  &=
    \frac{  }{} \, 
    \frac{2^{d-4} \Gamma \left(1-\frac{d}{2}\right) a^{d-2} b^{d-2} c^{2-d}(a+b)^{4-d} \,
   _2F_1\left(1,\frac{d-1}{2};d-1;\frac{4 a b}{(a+b)^2-c^2}\right)}{\Gamma \left(2-\frac{d}{2}\right)
   \left((a+b)^2-c^2\right)} +
   \cr &
   -\frac{  2^{d-4}\pi\,  b^{d-2} (a+b)^{4-d} \,
   _2F_1\left(\frac{1}{2},1;\frac{d}{2};\frac{4 b^2 c^2}{\left(-a^2+b^2+c^2\right)^2}\right)}{\sin\left(\frac{\pi  d}{2}\right) \Gamma
   \left(2-\frac{d}{2}\right) \Gamma \left(\frac{d}{2}\right) \left(-a^2+b^2+c^2\right)}+\cr& 
   -\frac{  2^{d-4}\pi \, a^{d-2} (a+b)^{4-d} \,
   _2F_1\left(\frac{1}{2},1;\frac{d}{2};\frac{4 a^2 c^2}{\left(a^2-b^2+c^2\right)^2}\right)}{\sin \left(\frac{\pi  d}{2}\right) \Gamma
   \left(2-\frac{d}{2}\right) \Gamma \left(\frac{d}{2}\right) \left(a^2-b^2+c^2\right)}+
   \cr&
   -\frac{ 2^{d-4}\pi \,  a^{d-2} b^{d-2} c^{2-d}  (a+b)^{4-d} \,
   _2F_1\left(\frac{1}{2},1;\frac{d}{2};\frac{4 a^2 b^2}{\left(a^2+b^2-c^2\right)^2}\right)}{ \sin \left(\frac{\pi  d}{2}\right)\Gamma
   \left(2-\frac{d}{2}\right) \Gamma \left(\frac{d}{2}\right) \left(a^2+b^2-c^2\right)}.
\end{align}
\end{lemma}
Explicit nontrivial formulae for the Appell functions are rare; it is another good point of our method its ability to produce such formulae.

 Similarly, by comparing Eqs (\ref{AAA}) and (\ref{bella formula}) we get three more summation formulae for the Appell series $F_4$:
\begin{align}
 & F_4\left(3-d,2-\frac d2,2-\frac d2,2-\frac d2,\frac{a^2}{c^2},\frac{b^2}{c^2}\right) = c^{6-2 d} \left(a^4+b^4+c^4-2 a^2 b^2-2 a^2 c^2-2 b^2
   c^2\right)^{\frac{d-3}{2}}  \cr & F_4\left(1,2-\frac d2,2-\frac d2,\frac d2,\frac{a^2}{c^2},\frac{b^2}{c^2}\right)= -\frac{c^2 \,
   }{a^2-b^2-c
   ^2} \   {}_2F_1\left(\frac{1}{2},1;\frac{d}{2};\frac{4 b^2
   c^2}{\left(-a^2+b^2+c^2\right)^2}\right)\cr&
   F_4\left(1,\frac d2, \frac d2,\frac d2,\frac{a^2}{c^2},\frac{b^2}{c^2}\right)= -\frac{c^2 \,
  }{a^2+b^2-c^2
}{}\  _2F_1\left(\frac{1}{2},1;\frac{d}{2};\frac{4 a^2
   b^2}{\left(a^2+b^2-c^2\right)^2}\right)
\end{align}

\vspace{3cm}

\noindent{\bf\large Acknowledgments. }  We thank Thibault Damour for many useful discussions and clarifications. SC and UM thank the members and the staff of IHES for their warm hospitality and support.
We also thank Paul McFadden, Adam Bzowski, and Kostas Skenderis as well as Alexei Morozov for valuable correspondence. We thank an anonymous referee for several comments that helped us to improve the presentation of our results.

\end{document}

\documentclass{article}
\usepackage{graphicx} 

\title{loopflat}
\author{ugo.moschella }
\date{February 2023}

\begin{document}

\maketitle

\section{Introduction}


\begin{thebibliography}{99}

\bibitem{Groote:2005ay}
S.~Groote, J.~G.~Korner and A.~A.~Pivovarov,
``On the evaluation of a certain class of Feynman diagrams in x-space: Sunrise-type topologies at any loop order,''
Annals Phys. \textbf{322} (2007), 2374-2445


\bibitem{Bzowski:2013sza}
A.~Bzowski, P.~McFadden and K.~Skenderis,
``Implications of conformal invariance in momentum space,''
JHEP \textbf{03} (2014), 111

\bibitem{Bzowski:2015yxv}
A.~Bzowski, P.~McFadden and K.~Skenderis,
``Evaluation of conformal integrals,''
JHEP \textbf{02} (2016), 068

\bibitem{Farrow:2018yni}
J.~A.~Farrow, A.~E.~Lipstein and P.~McFadden,
``Double copy structure of CFT correlators,''
JHEP \textbf{02} (2019), 130

\bibitem{Lipstein:2019mpu}
A.~E.~Lipstein and P.~McFadden,
``Double copy structure and the flat space limit of conformal correlators in even dimensions,''
Phys. Rev. D \textbf{101} (2020) no.12, 125006

\bibitem{Mishnyakov:2023wpd}
V.~Mishnyakov, A.~Morozov and P.~Suprun,
``Position Space Equations for Banana Feynman Diagrams,''
[arXiv:2303.08851 [hep-th]].

\bibitem{Bloch:1937pw}
F.~Bloch and A.~Nordsieck,
``Note on the Radiation Field of the electron,''
Phys. Rev. \textbf{52} (1937), 54-59
doi:10.1103/PhysRev.52.54

 \bibitem{Yennie}
D.~R.~Yennie, S.~C.~Frautschi and H.~Suura,
``The infrared divergence phenomena and high-energy processes,''
Annals Phys. \textbf{13} (1961), 379-452

\bibitem{Weinberg:infrared}
S.~Weinberg,
``Infrared photons and gravitons,''
Phys. Rev. \textbf{140} (1965), B516-B524

\bibitem{Arkani-Hamed:2013jha}
N.~Arkani-Hamed and J.~Trnka,
``The Amplituhedron,''
JHEP \textbf{10} (2014), 030

\bibitem{Arkani-Hamed:2014dca}
N.~Arkani-Hamed, A.~Hodges and J.~Trnka,
``Positive Amplitudes In The Amplituhedron,''
JHEP \textbf{08} (2015), 030

\bibitem{Arkani-Hamed:2020blm}
N.~Arkani-Hamed, T.~C.~Huang and Y.~T.~Huang,
``The EFT-Hedron,''
JHEP \textbf{05} (2021), 259

\bibitem{Arkani-Hamed:2022cqe}
N.~Arkani-Hamed, A.~Hillman and S.~Mizera,
``Feynman polytopes and the tropical geometry of UV and IR divergences,'' 
Phys. Rev. D \textbf{105} (2022) no.12, 125013

\bibitem{Chetyrkin:1981qh}
K.~G. Chetyrkin, and F.~V. Tkachov,
``Integration by Parts: The  Algorithm to Calculate beta Functions in 4 Loops,''
Nucl. Phys. \textbf{B}192 (1981) {159--204}

\bibitem{Laporta:2003jz}
S. Laporta,
``Calculation of Feynman integrals by difference equations,''
Acta Phys. Polon. \textbf{B}{34}, (2003) {5323--5334}

\bibitem{AomotoKita}
K. Aomoto and M. Kita, ``Theory of Hypergeometric Functions,'' Springer Monographs in Mathematics (Springer Japan, 2011).

\bibitem{MatsuGoto}
Y. Goto and K. Matsumoto, ``The monodromy representation and twisted period relations for Appell's hypergeometric function F4,'' Nagoya Math. J. 217 (2015) 61--94.

\bibitem{MatsuTaka}
S.-J. Matsubara-Heo and N. Takayama, ``An algorithm of computing cohomology intersection number of hypergeometric integrals,'' Nagoya Math. J. 1--17, DOI: 10.1017/nmj.2021.2 (2019)

\bibitem{Matsumoto}
K. Matsumoto, ``Introduction to the Intersection Theory for Twisted Homology and Cohomology Groups,'' PoS MA2019, 007 (2022).

\bibitem{MastroMiz}
P. Mastrolia and S. Mizera, ``Feynman Integrals and Intersection Theory,'' JHEP \textbf{02} (2019), 139

\bibitem{Frell1}
H. Frellesvig, {\em et al.} ``Decomposition of Feynman Integrals on the Maximal Cut by Intersection Numbers,'' JHEP \textbf{05} (2019), 153

\bibitem{Frell2}
H. Frellesvig, {\em et al.} ``Vector Space of Feynman Integrals and Multivariate Intersection Numbers,'' Phys. Rev. Lett. 123 (2019), 201602

\bibitem{Frell3}
H. Frellesvig, {\em et al.} ``Decomposition of Feynman Integrals by Multivariate Intersection Numbers,'' JHEP \textbf{03} (2021), 027

\bibitem{Chestnov}
V. Chestnov, {\em et al.} ``Macaulay Matrix for Feynman Integrals: Linear Relations and Intersection Numbers,'' arxiv: hep-th/2204.12983.

\bibitem{Mizera1}
S. Mizera, ``Scattering Amplitudes from Intersection Theory,'' Phys. Rev. Lett. 120 (2018), 141602

\bibitem{Mizera2}
S. Mizera, ``Status of Intersection Theory and Feynman Integrals,'' PoS MA2019, 016 (2019)

\bibitem{Weinzierl}
S. Weinzierl, ``Feynman Integrals,'' arxiv: hep-th:2201.03593.

\bibitem{Mastrolia}
P. Mastrolia, ``From Diagrammar to Diagrammalgebra,'' PoS MA2019, 015

\bibitem{MandalGasp}
M.~K. Mandal and F. Gasparotto, ``On the Application of Intersection Theory to Feynman Integrals: the multivariate case,''
PoS MA2019, 019

\bibitem{CaronPokr1}
S. Caron-Huot and A. Pokraka, ``Duals of Feynman integrals. Part I. Differential equations,'' JHEP \textbf{12} (2021), 045

\bibitem{CaronPokr2}
S. Caron-Huot and A. Pokraka, ``Duals of Feynman Integrals. Part II. Generalized unitarity. JHEP \textbf{04} (2022), 078

\bibitem{CheFreGaMaMa}
V. Chestnov, H. Frellesvig, F. Gasparotto, F., M.~K. Mandal and P. Mastrolia, ``Intersection Numbers from Higher-order Partial
Differential Equations,'' arxiv: hep-th/2209.01997

\bibitem{Cacciatori:2022mbi}
S.~L.~Cacciatori and P.~Mastrolia,
``Intersection Numbers in Quantum Mechanics and Field Theory,''
arXiv:2211.03729 [hep-th]

\bibitem{Bloch:2014qca}
S.~Bloch, M.~Kerr and P.~Vanhove,
``A Feynman integral via higher normal functions,''
Compos. Math. \textbf{151} (2015) no.12, 2329-2375

\bibitem{Bonisch:2021yfw}
K.~B\"onisch, C.~Duhr, F.~Fischbach, A.~Klemm and C.~Nega,
``Feynman integrals in dimensional regularization and extensions of Calabi-Yau motives,'' 
JHEP \textbf{09} (2022), 156

\bibitem{Duhr:2022dxb}
C.~Duhr, A.~Klemm, C.~Nega and L.~Tancredi,
``The ice cone family and iterated integrals for Calabi-Yau varieties,''
arXiv:2212.09550 [hep-th]

\bibitem{Pogel:2022vat}
S.~P\"ogel, X.~Wang and S.~Weinzierl,
``Bananas of equal mass: any loop, any order in the dimensional regularisation parameter,'' 
arXiv:2212.08908 [hep-th]

\bibitem{Broadhurst1}
D.~Broadhurst and D.~P.~Roberts,
``Quadratic relations between Feynman integrals,''
PoS \textbf{LL2018} (2018), 053

\bibitem{Broadhurst2}
D.~Broadhurst and A.~Mellit,
``Perturbative quantum field theory informs algebraic geometry,''
PoS \textbf{LL2016} (2016), 079

\bibitem{Zhou1}
Y.~Zhou,
``Some algebraic and arithmetic properties of Feynman diagrams,''
doi:10.1007/978-3-030-04480-0\_19

\bibitem{Zhou2}
Y.~Zhou,
``Wro\'nskian factorizations and Broadhurst\textendash{}Mellit determinant formulae,''
Commun. Num. Theor. Phys. \textbf{12} (2018), 355-407

\bibitem{Broadhurst3}
D.~Broadhurst,
``Feynman integrals, L-series and Kloosterman moments,''
Commun. Num. Theor. Phys. \textbf{10} (2016), 527-569

\bibitem{Zhou3}
Y.~Zhou,
``Wro\'nskian algebra and Broadhurst\textendash{}Roberts quadratic relations,''
Commun. Num. Theor. Phys. \textbf{15} (2021) no.4, 651-741

\bibitem{Adams}
L.~Adams, C.~Bogner and S.~Weinzierl,
``The two-loop sunrise integral around four space-time dimensions and generalisations of the Clausen and Glaisher functions towards the elliptic case,''
J. Math. Phys. \textbf{56} (2015) no.7, 072303

\bibitem{Remiddi}
E.~Remiddi and L.~Tancredi,
``Schouten identities for Feynman graph amplitudes; The Master Integrals for the two-loop massive sunrise graph,''
Nucl. Phys. B \textbf{880} (2014), 343-377

\bibitem{Zagier}
D. Zagier, ``The Dilogarithm Function.'' In: Cartier, P., Moussa, P., Julia, B., Vanhove, P. (eds) Frontiers in Number Theory, Physics, and Geometry II (2007) Springer, Berlin, Heidelberg.

\bibitem{Vanhove}
S.~Bloch, P.~Vanhove,
``The elliptic dilogarithm for the sunset graph,''
Journal of Number Theory, Volume 148 (2015) 328-364,

\bibitem{Ford:1992pn}
C.~Ford, I.~Jack and D.~R.~T.~Jones,
``The Standard model effective potential at two loops,''
Nucl. Phys. B \textbf{387} (1992), 373-390
[erratum: Nucl. Phys. B \textbf{504} (1997), 551-552]

\bibitem{bailey} 
W. N. Bailey. Some infinite integrals involving Bessel functions, Proc. London Math. Soc. (2), 40,  37-48 (1936).

\bibitem{bailey2} 
W. N. Bailey, Some infinite integrals involving Bessel functions II, J. Lond. Math. Soc. s1-11, 16-20 (1936).

\bibitem{batemanI} Harry  Bateman, Higher Transcendental Functions, Vol.I. New York:  McGraw-Hill Book Company  (1953), 

\bibitem{bateman2} Harry  Bateman, Higher Transcendental Functions, Vol.II. New York:  McGraw-Hill Book Company  (1953), 

\bibitem{abramowitz} 
Handbook of Mathematical Functions with Formulas, Graphs, and Mathematical Tables,
M. Abramowitz, I. A. Stegun, editors, tenth edition (1972)

\end{thebibliography}
\end{document}
{\color{blue}
With the change of variable $s=\sqrt{1-t}$, we have
\begin{align}
    I'_{z,1}(0)=-\frac 12  \int_0^1 \frac {dt}{\sqrt{1-t}} (1-zt)^{k-2} \log(1-zt)=-\int_0^1 \frac {ds}{1-z+zs^2} \log(1-z+zs^2)\\
    =-\frac 1z \int_0^1 \frac {ds}{\frac {1-z}z+s^2} (\log z+ \log (\frac {1-z}z+s^2))\\
    =-\frac {\log z}z \int_0^1 \frac {ds}{\frac {1-z}z+s^2}-\frac 1z \int_0^1 \frac {ds}{\frac {1-z}z+s^2} \log (\frac {1-z}z+s^2).
\end{align}
Now set $s=u \sqrt {\frac {1-z}z}$:
\begin{align}
    I'_{z,1}(0)=-\frac {\log (1-z)}{\sqrt {z(1-z)}} \int_0^{\sqrt{\frac{z}{1-z}}} \frac {du}{1+u^2}-\frac {1}{\sqrt {z(1-z)}} \int_0^{\sqrt{\frac{z}{1-z}}} 
    \frac {du}{1+u^2} \log (1+u^2)\\
    =-\frac {\log (1-z)}{\sqrt {z(1-z)}} \arctan \sqrt {\frac z{1-z}}-\frac {1}{\sqrt {z(1-z)}} \int_0^{\sqrt{\frac{z}{1-z}}} 
    \frac {du}{1+u^2} \log (1+u^2).
\end{align}
We write
\begin{align}
    \int_0^{\sqrt{\frac{z}{1-z}}} \frac {du}{1+u^2} \log (1+u^2)=\frac 12 \int_0^{\sqrt{\frac{z}{1-z}}} du \left( \frac 1{1+iu} +\frac 1{1-iu} \right)
    \left( \log (1+iu) + \log (1-iu)\right)\\
    =\frac 12 \int_0^{\sqrt{\frac{z}{1-z}}} du \frac 1{1+iu} \left( \log (1+iu) + \log (1-iu) \right)+c.c.\\
    =\frac 1{2} \left(\frac 1{2i}\log^2 (1+i\sqrt{\frac{z}{1-z}}) +\int_0^{\sqrt{\frac{z}{1-z}}} du \frac 1{1+iu} \log (1-iu) \right)+c.c.\\
    =\frac 1{2} \log \frac 1{1-z} \frac 1{2i}\log \frac {1+i\sqrt{\frac{z}{1-z}}}{1-i\sqrt{\frac{z}{1-z}}}+\left(\frac 1{2}\int_0^{\sqrt{\frac{z}{1-z}}} du \frac 1{1+iu} \log (1-iu) +c.c. \right)\\
    =\frac 1{2} \log \frac 1{1-z} \arctan \sqrt{\frac{z}{1-z}}+\left(\frac 1{2i}\int_{\frac 12}^{\frac 12+\frac i2\sqrt{\frac{z}{1-z}}} dx \frac 1{x} \log (2-2x) +c.c. \right)\\
    =\frac 1{2} \log \frac 1{1-z} \arctan \sqrt{\frac{z}{1-z}}+\frac 1{2i} \log 2 \log \frac {1+i\sqrt{\frac{z}{1-z}}}{1-i\sqrt{\frac{z}{1-z}}} 
    +\frac 1{2i} \left(\int_{\frac 12}^{\frac 12+\frac i2\sqrt{\frac{z}{1-z}}} dx \frac {\log (1-x)}x-c.c. \right)\\
    =\log \frac 2{\sqrt{1-z}} \arctan \sqrt{\frac{z}{1-z}}
    -\frac 1{2i} \left( {\rm Li}_2 (\frac 12+\frac i2\sqrt{\frac{z}{1-z}})-{\rm Li}_2 (\frac 12-\frac i2\sqrt{\frac{z}{1-z}}) \right)
\end{align}
where we used $x=(1+iu)/2$. Therefore, we get}

\subsubsection{Case $\j=2k$, $k\geq 2$}
For $k\geq 2$ the ingredients are
\begin{align}
    \frac 1{\sin^2(\pi\varepsilon)}=\frac 1{\pi^2\varepsilon^2}+\frac 13+O(\varepsilon^2),
\end{align}
We also need
\begin{align}
    \Gamma(2-2k-2\varepsilon) \left(\frac {-S(m_1,m_2,m_3)}{16\pi^2}\right)^{-\varepsilon}=&
    \frac {\left(\frac 1{2\varepsilon} -\gamma+H_{2k-2}-\frac 12 \log \frac {-S(m_1,m_2,m_3)}{16\pi^2} \right)+\ldots,}{(2k-2)!}\cr
    \left( \frac {m_1m_2}{4\pi}\right)^{-2\varepsilon}=& 1-2\varepsilon \log \frac {m_1m_2}{4\pi}+2\varepsilon^2 \log^2 \frac {m_1m_2}{4\pi}+\ldots,\cr
    \frac 1{\Gamma(k-\varepsilon)}=& \frac 1{(k-1)!}\left( 1+\psi(k)\varepsilon +(\psi'(k)-\psi(k)^2) \frac {\varepsilon^2}2 \right)+\ldots,
\end{align}
and 
we recall 
\begin{align}
    \psi(k)=&-\gamma+H_{k-1}, \qquad H_{n}=\sum_{j=1}^{n-1}\frac 1j,\\
    \psi'(k)=& \frac {\pi^2}6-H_{n-1,2}, \qquad H_{n,2}=\sum_{j=1}^{n-1}\frac 1{j^2}.
\end{align}
Finally, we need the expansion of $I_{z,k}(\varepsilon)\equiv{}_{2}F_1(1,2-k+\varepsilon;\frac 32;z)$, up to order 2 for $k>1$ and up to order 1 for $k=1$.
Using 
\begin{align}
    I_{z,k}({\varepsilon})=\frac 12 \int_0^1 \frac {dt}{\sqrt{1-t}} \frac 1{(1-zt)^{2-k+\varepsilon}}, 
\end{align}
we get
\begin{align}
    I'_{z,k}(0)=& -\frac 12  \int_0^1 \frac {dt}{\sqrt{1-t}} (1-zt)^{k-2} \log(1-zt), \\
    I''_{z,k}(0)=& \frac 12  \int_0^1 \frac {dt}{\sqrt{1-t}} (1-zt)^{k-2} \log^2(1-zt).
\end{align}
A quite involved but direct computation gives

\begin{align*}
    I'_{z,1}(0)=&-\frac {\log (2\sqrt {1-z})}{\sqrt {z(1-z)}} \arctan \sqrt {\frac z{1-z}}\\
    &+\frac 1{2i\sqrt{z(1-z)}}\left({\rm Li}_2 \left( \frac 12+\frac i2\sqrt{\frac{z}{1-z}}\right)-{\rm Li}_2 \left( \frac 12-\frac i2\sqrt{\frac{z}{1-z}} \right)\right).
\end{align*}

\begin{align}
    I'_{z,1}(0)=& -\frac{1}{\sqrt{z(1-z)}} \arctan{\sqrt \frac{z}{1-z}}\log (2\sqrt{1-z})\cr
    &+\frac 1{2i\sqrt{z(1-z)}}\left({\rm Li}_2 \left( \frac 12+\frac i2\sqrt{\frac{z}{1-z}}\right)-{\rm Li}_2 \left( \frac 12-\frac i2\sqrt{\frac{z}{1-z}} \right)\right), \\
    I'_{z,2}(0)=& 2-2 \frac {\sqrt {1-z}}{\sqrt z} \arctan \frac {\sqrt z}{\sqrt {1-z}}; \\
    I''_{z,2}(0)=& 4i\sqrt{\frac{1-z}{z}}\left({\rm Li}_2 \left( \frac {1+i\sqrt{\frac{1-z}{z}}}{1-i\sqrt{\frac{1-z}{z}}}\right)+\frac {\pi^2}{12}\right)
    -4i\sqrt{\frac{1-z}{z}}\arctan^2 \sqrt{\frac{z}{1-z}}\cr
    &+8-8\sqrt{\frac{1-z}{z}}\arctan \sqrt{\frac{z}{1-z}} +4\sqrt{\frac{1-z}{z}} \arctan \sqrt{\frac{z}{1-z}}\log (4-4z).
\end{align}
Putting everything together there follows an $\varepsilon$-expansion for all even dimensions. Let us write it down for the cases of $d=2$ and $d=4$.
For $k=1$ we can use 
\begin{align}
    {}_2F_1\left(1,1;\frac 32;z^2\right) =\frac 1{z(1-z^2)^{\frac 12}} \arcsin{z},
\end{align}
to write 
\begin{align}
 &  I(m_1,m_2,m_3,2-2\varepsilon)=\cr &=  \frac 1{\varepsilon} \left( -\frac 1{16 \pi (-S)^\frac 12} +\left[\frac {1}{16 \pi^2m_1 m_2} 
   \frac 1{(1-z_{1,2,3}^2)^{\frac 12}} \arcsin{z_{1,2,3}} +cyc\{1,2,3\} \right] \right)
   \cr
   &+\frac 1{8\pi (-S)^\frac 12} \left(\gamma +\frac 12 \log \frac {-S}{16\pi^2} \right)\cr
   &-\left[\frac {m_1^2+m_2^2-m_3^2}{32\pi^2 m_1^2 m_2^2} \left(2(\gamma +\log \frac {m_1m_2}{4\pi})\frac 1{z_{1,2,3}(1-z_{1,2,3}^2)^{\frac 12}} \arcsin{z_{1,2,3}} \right.
   \right. \cr
   & +\frac 2{\sqrt {1-z^2_{1,2,3}}} Cl_2\left(2\arctan \frac {\sqrt{1-z^2_{1,2,3}}}{z_{1,2,3}}\right)\cr
   & \left.\left.-\frac 1{\sqrt {1-z_{1,2,3}}} \left(\pi-2\arctan \frac {\sqrt{1-z^2_{1,2,3}}}{z_{1,2,3}}\right) \log (4-4z^2_{1,2,3})\right)
   +cyc\{1,2,3\} \right] 
    +O(\varepsilon),\cr &
\end{align} 
where   $S = S(m_1,m_2,m_3)$ and  we introduced
\begin{align}
    z_{1,2,3}=\frac {m_1^2+m_2^2-m_3^2}{2m_1m_2},
\end{align}
and $+cyc\{1,2,3\}$ obviously means that one has to add the terms obtained by cyclic permutations of all the indices $1,2,3$.
Now, we can use that
\begin{align}
 \arctan \frac {\sqrt{1-z^2_{1,2,3}}}{z_{1,2,3}}=\arccos z_{123}.
\end{align}
Moreover, the term in square parenthesis in the above formula can be rewritten as
\begin{align}
 \frac 1{8\pi^2 (-S)^\frac 12}  \arcsin{z_{1,2,3}} +cyc\{1,2,3\}.
\end{align}
Now, one easily checks that
\begin{align}
 \frac {\partial}{\partial m_j} (\arcsin{z_{1,2,3}} +cyc\{1,2,3\})=0.
\end{align}
Therefore, it is constant and can be evaluated by considering $m_j=m$, which gives $(\arcsin{z_{1,2,3}} +cyc\{1,2,3\})=\frac \pi2$. The same trick shows that the coefficients of $\gamma$ in the above expression cancel out and we finally get
\begin{align}
 &  I(m_1,m_2,m_3,2-2\varepsilon)= \frac 1{16\pi (-S)^\frac 12} \log \left(\frac {-S}{(m_1m_2m_3)^{\frac 43}}\right) \cr
 &-\left[\frac 1{12\pi^2 (-S)^\frac 12} \log \left(\frac {m_1m_2}{m_3^2}\right) \arcsin{z_{1,2,3}} 
   \right. \cr
   & +\frac {1}{16\pi^2 m_1 m_2} \left(\frac {2z_{1,2,3}}{\sqrt {1-z^2_{1,2,3}}} Cl_2\left(2\arccos z_{123}\right) \right.\cr
   & \left.\left.-\frac {z_{1,2,3}}{\sqrt {1-z_{1,2,3}}} \left(\pi-2\arccos z_{123}\right) \log (4-4z^2_{1,2,3})\right)
   +cyc\{1,2,3\} \right] 
    +O(\varepsilon),&
\end{align} 
which, as expected, is finite for $\varepsilon\to 0$. Also, notice that the arguments of the logarithms are adimensional, showing that the result is independent of the eventual choice of an extra energy scale $\mu$. \\
{\color{blue} Questa formula dovrebbe coincidere con 
\begin{align*}
 I(m_1,m_2,m_3,2)=2\pi^2 \frac {D(m_1,m_2,m_3)}{\sqrt {-S}},
\end{align*}
dove D \`e il dilogaritmo di Bloch-Wigner, vedi la formula (3.10) e appendice A in\\ https://arxiv.org/pdf/1309.5865.pdf.\\
Penso che con gli stessi trucchi sopra si possa semplificare un pochino anche la formula qui sotto per d=4. }